\documentclass[twocolumn,floats,pra,aps]{revtex4}

\usepackage{epsfig,psfrag,subfigure}
\usepackage{graphicx}
\usepackage{amsfonts}
\usepackage{amsmath}
\usepackage{amssymb}
\usepackage{exscale}
\usepackage{color}
\usepackage{subeqnarray}
\def\lan{\langle}
\def\ran{\rangle}
\def\va{\varepsilon}
\def\dag{\dagger}

\def\vk{{\bf k}}
\def\vK{{\bf K}}

\def\vP{{\bf P}}
\def\vp{{\bf p}}
\def\vq{{\bf q}}

\def\vQ{{\bf Q}}

\def\v0{{\bf 0}}
\newcommand{\bd}{\begin{equation}}
\newcommand{\ed}{\end{equation}}
\newcommand{\be}{\begin{equation}}
\newcommand{\ee}{\end{equation}}
\newcommand{\bt}{\begin{split}}
\newcommand{\et}{\end{split}}

\newcommand{\bn}{\begin{align}}
\newcommand{\en}{\end{align}}
\newcommand{\bea}{\begin{eqnarray}}
\newcommand{\eea}{\end{eqnarray}}
\newcommand{\ba}{\begin{array}}
\newcommand{\ea}{\end{array}}
\newcommand{\nn}{\nonumber}

\begin{document}


\title{Coboson many-body formalism for atom-dimer scattering length}

\author{Shiue-Yuan Shiau}
\affiliation{Physics Division, National Center for Theoretical Sciences, Hsinchu 30013, Taiwan}
\affiliation{Research Center for Applied Sciences, Academia Sinica, Taipei, 115 Taiwan}
 \author{Ching-Hang Chien} 
 \affiliation{Research Center for Applied Sciences, Academia Sinica, Taipei, 115 Taiwan}
 \author{Yia-Chung Chang} 
 \affiliation{Research Center for Applied Sciences, Academia Sinica, Taipei, 115 Taiwan}
 \author{Monique Combescot}
\affiliation{Sorbonne Universit\'e, CNRS, Institut des NanoSciences de Paris, 75005 Paris, France}



\begin{abstract}

We use the composite boson (coboson) many-body formalism to tackle scattering lengths for cold fermionic atoms. We show that bound dimers can be taken as elementary entities provided that fermion exchanges between them are treated exactly, as can be done through the coboson formalism. This alternative tool extended to cold atom physics  not only makes transparent many-body processes  through Shiva diagrams specific to cobosons, but also simplifies calculations. 
Indeed, the integral equation we derive for the atom-dimer scattering length and  solve by restricting  the dimer relative motion  to the ground state, gives values in remarkable agreement with the exact scattering length values for all fermion mass ratios. This remarkable agreement also holds  true for the dimer-dimer scattering length, except for equal fermion masses where our restricted procedure gives a value slightly larger than the accepted one ($0.64a_d$ instead of $0.60a_d$). All this proves that the scattering of a cold-atom dimer with an atom or another dimer is essentially controlled by the dimer relative-motion ground  state, a physical result not obvious at first.   

\end{abstract}


\maketitle

\section{Introduction}

Interest in few-body systems started long before the spectacular development of cold atom physics. In semiconductors, Lambert has predicted\cite{Lampert} in the 1960's that  semiconductor electrons and holes can form multi-body complexes such as excitons, trions, and biexcitons. In nuclear physics, the hadron-hadron scattering has also been an important subject of study for decades. Another peculiar phenomenon on its own right is the  universal behavior that occurs in three-particle systems initiated by Efimov\cite{Efimov,Naidon}, now known as  Efimov physics.

The scattering  of three fermions $(\alpha,\alpha,\beta)$ interacting through a short-range attractive potential between different-species fermions only, is also  known  to exhibit universality. Indeed, the atom-dimer scattering length for equal fermion masses is equal to $1.18a_d$ where $a_d$ is the scattering length of the $(\alpha,\beta)$ fermion pair, whatever the interaction strength\cite{Skorniakov}. This problem has been taken up again in the 2000's and extended to arbitrary mass ratios due to its interest in cold atom systems,  by directly solving the three-body Schr\"{o}dinger equation\cite{Petrov2003,Petrov2005}, through a finite-volume lattice method\cite{Bour2012,Elhatisari}, through the effective field theory\cite{Bedaque,Hammer}, or through the field theoretical many-body procedure\cite{Iskin2008,Iskin2010,Levisen2011}. A precise analysis of  limiting mass ratios can  be found in \cite{Alzettopra2010}. \

In this work, we reconsider the three-fermion scattering problem from a different perspective using the coboson many-body formalism\cite{MoniqPhysreport,book}. This formalism deals from the very first lines with cobosons. Calculation of many-body effects between cobosons or between cobosons and other particles relies on the commutation relations between cobosons and the system Hamiltonian. It fundamentally differs in both technique and language from the effective field theory and the field theoretical many-body procedure. We have  shown\cite{Moniq2016pra} that the coboson formalism takes a much simpler form for particles like fermionic atoms which interact between different fermion species only. Besides its simplicity, the great advantage of this rather new many-body formalism is to provide a direct, visual  understanding of the role of the Pauli exclusion principle in few-fermion scattering problems, through Shiva diagrams that are specific to cobosons, as shown in Figs.~\ref{fig:2} and \ref{fig:3}. Shiva diagrams allow visualizing in a transparent way the interplay between energy-like interactions, dimensionless fermion exchanges and the Pauli exclusion principle between the coboson constituents. More details on how Shiva diagrams are used to calculate physical quantities can be found in \cite{book}.  The important role of the Pauli exclusion principle  in few-body scatterings   deserves some discussions, as we will do here. \

The main result of this work is that the integral equation for the atom-dimer scattering length derived from the coboson many-body formalism leads to accurate values for {\it all} fermion mass ratios, despite the fact that we numerically solve it by restricting the dimer relative-motion states to the  ground state. Furthermore, the integral equation becomes analytically exact both in the large and small mass ratio limits. In this work, we also reconsider dimer-dimer scattering. For this case too, the integral equation restricted to the relative-motion ground state becomes analytically exact in the large mass ratio limit, so is the scattering length derived from it,  while for equal masses the numerically obtained scattering length   is found to be equal to $0.64a_d$\cite{Shiau2016pra}, which  is slightly larger than the commonly accepted value\cite{Levisen2011,Petrov2004,Petrov2005,Brodsky2006,Alzetto2013,Incao}, $0.60a_d$. \

The above  analytical and numerical results show that cold atom  dimers can be treated as elementary  entities in their many-body interactions with atoms or other dimers, provided  the fermion exchanges induced by the Pauli exclusion principle are handled exactly. This indicates that,  as far as $s$-wave scattering is concerned, the dimer relative-motion excited states play little role in the interaction of a dimer with other dimer or atom, and a  small role in the interaction of two dimers for equal masses. We attribute this effect to the short-rangeness of the atom-atom potential. The cold-atom system we here consider is in stark contrast with Coulombian system where a particle can impinge over a large distance on another particle  because of the finite-rangeness of the Coulomb potential to create partial waves with all possible angular momenta.  \

This paper is organized as follows. In Sec.~\ref{sec2:sd}, we consider a single dimer made of $(\alpha, \beta)$ atoms, and we determine the threshold of the potential amplitude for the dimer to have a bound state. In Sec.~\ref{sec3:ad}, we derive the atom-dimer scattering length. The coboson formalism allows us to link the scattering length to the energy induced by the atom-dimer interaction, while visualizing the fermion exchanges between atom and dimer in the scattering processes. We then analyze the  integral equation in two  mass ratio limits, and compute the scattering length for all mass ratios. In Sec.~\ref{sec4:dd}, we analyze the dimer-dimer scattering in a similar way. We then conclude.

\section{Dimer scattering length\label{sec2:sd}}

\begin{figure}[t!]
\begin{center}
\includegraphics[trim=7.5cm 8.3cm 7cm 6cm,clip,width=2.8in]{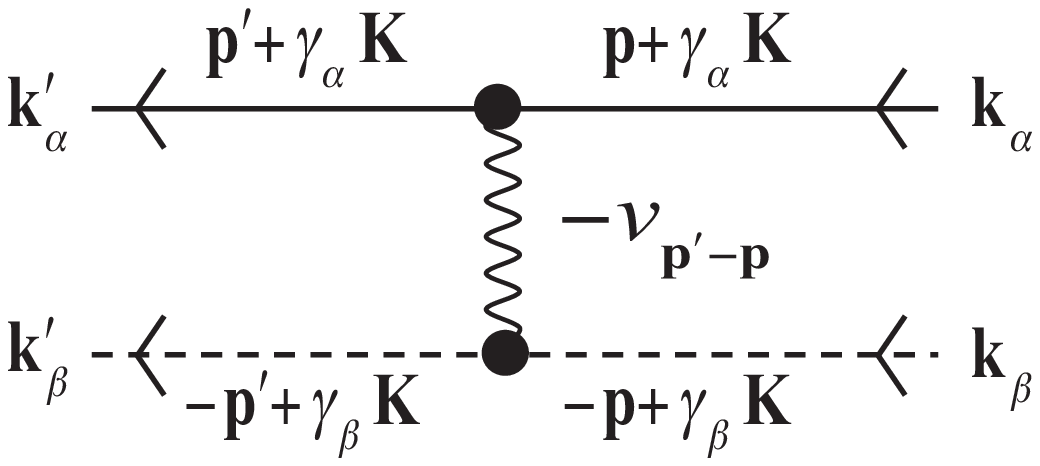}
\caption{\small Diagrammatic representation of the potential given in Eq.~(\ref{eq:VintofBKp}) acting between $\alpha$ and $\beta$ fermionic atoms. \label{fig:0}}
\end{center}
\end{figure}

We consider two species of fermionic atoms, $\alpha$ and $\beta$. The  kinetic part of the Hamiltonian reads $H_0=\sum_{\vk}\va_\vk^{(\alpha) }a^\dag_\vk a_\vk+\sum_{\vk}\va_\vk^{(\beta)}b^\dag_\vk b_\vk $ with $\va_\vk^{(\alpha,\beta)}=\vk^2/2m_{\alpha, \beta}$, the operators $a^\dag_\vk$ and $b^\dag_\vk$ denoting  creation operators for free atoms $\alpha$ and $\beta$ with momentum $\vk$, respectively. Because of  low energy and temperature,  same-species atoms  do not interact while different-species atoms  interact via a short-ranged attractive potential that can be written as
\be
V=
-\sum_{\vK,\vp,\vp'} B^\dag_{\vK,\vp'} v_{\vp'-\vp}B_{\vK, \vp}\, \label{eq:VintofBKp}
\ee
with $B^\dag_{\vK,\vp}=a^\dag_{\vp+\gamma_\alpha \vK}b^\dag_{-\vp+\gamma_\beta \vK}$ (see Fig.~\ref{fig:0}). 
For $\gamma_\alpha=1-\gamma_\beta= m_\alpha/(m_\alpha+m_\beta)$, the operator $B^\dag_{\vK,\vp}$ creates a free fermion pair with center-of-mass momentum  $\vK$ that remains constant in the scattering, and relative-motion momentum $\vp$ that changes from $\vp$ to $\vp'$ with a scattering amplitude $(-v_{\vp'-\vp})$. \

The coboson many-body formalism treats the single-dimer eigenstates as entities while  handling fermion exchanges between them in an exact way. This formalism reads in terms of dimer creation operators $B^\dag_i$ defined as $|i\ran=B^\dag_i|v\ran$ with $(H-E_i)|i\ran=0$ where $|v\ran$ denotes the vacuum state, the Hamiltonian $H$ is equal to $H_0+V$, and $E_i$ is the dimer $i$ energy. As the potential $V$ conserves the dimer center-of-mass momentum, it is convenient to have the center-of-mass momentum $\vK_i$ of dimer $i$ appearing explicitly in its creation operator, that is, to write $B^\dag_i$ as  
\be
B^\dag_i\equiv B^\dag_{\vK_i,\nu_i}=\sum_{\vp}   B^\dag_{\vK_i,\vp}\lan \vp|\nu_i\ran \,.\label{BiBvKinu1} 
\ee
We then note that $VB^\dag_{\vK,\vp}|v\ran=-\sum_{\vp^\prime}v_{\vp^\prime-\vp}B^\dag_{\vK,\vp^\prime}|v\ran$, while $\big[H_0,B^\dag_{\vK,\vp}\big]_-=E_ {\vK,\vp}  B^\dag_{\vK,\vp}$ with   $E_ {\vK,\vp}=\va_{\vK}^{(d)}+\va_{\vp}$ where $\va_{\vK}^{(d)}=\vK^2/2(m_\alpha+m_\beta)$ and $\va_{\vp}=\vp^2/2\mu_d$ for  $\mu_d^{-1}=m_\alpha^{-1}+m_\beta^{-1}$. As a result, by projecting the dimer Schr\"{o}dinger equation over a free pair state, we see that 
$\lan \vp|\nu_i\ran$ corresponds to the dimer relative-motion wave function in momentum space provided that
\be
0=\big(\va_{\vp}-\va_{\nu_i}\big)\lan \vp|\nu_i\ran-\sum_{\vp'}v_{\vp-\vp'} \lan \vp'|\nu_i\ran\,,\label{singlepaires}
\ee
the energy of  dimer $i$ being given by $E_i=\va_{\vK_i}^{(d)}+\va_{\nu_i}$. \

For   short-ranged atom-atom potential, $v_{\vp-\vp'}$ in Eq.~(\ref{eq:VintofBKp}) can be taken in a separable form, 
\be
v_{\vp-\vp'}=v w_\vp w_{\vp'}\,,
\ee where $v$ is a constant and $w_\vp =1$ up to a large cutoff $q_c$ and zero beyond. This cutoff, which mimics the natural decrease of the potential with momentum transfer, allows a proper handling of spurious divergences at large momenta. Equation (\ref{singlepaires}) then gives the single-dimer relative-motion eigenstates through
\be
\lan \vp|\nu_i\ran=\frac{v \,w_\vp}{\va_{\vp}-\va_{\nu_i}}\sum_{\vp'}w_{\vp'} \lan \vp'|\nu_i\ran\,,\label{singlepaires1}
\ee
which readily leads to an integral equation for the relative-motion energy 
\be
\frac{1}{v}=\sum_\vp \frac{w_\vp}{\va_{\vp}-\va_{\nu_i}}\, ,\label{sol_singlepair}
\ee 
since $w_\vp=w_\vp^2$. In three dimensions, this equation is known to allow one  bound state only, $\va_{\nu_g}<0$, provided that $v$ is larger than a $v^*$ threshold that corresponds to $\va_{\nu_g}=0$, namely, 
\be
\frac{1}{v^*}=\sum_\vp\frac{w_\vp}{\va_\vp}\,.\label{defv*}
\ee
Equation (\ref{sol_singlepair})  gives the energy, $\mathcal{E}_g^{(d)}\equiv E_g=\va_{\nu_g}$, of the dimer ground state denoted as $g=(\vK_g=\v0,\nu_g)$, through
\be
\frac{1}{v}-\frac{1}{v^*}=\sum_\vp w_\vp \frac{\mathcal{E}_g^{(d)}}{\va_{\vp}\big(\va_{\vp}-\mathcal{E}_g^{(d)}\big)}\, ,\label{7}
\ee
in which $w_\vp$ can be replaced by 1 since the $\vp$ sum now converges for large momentum. A simple way to solve this equation and relate the dimer ground state energy $ \mathcal{E}_g^{(d)}$ to the potential amplitude $v$ is to rewrite the dimer ground state energy as
\be
\mathcal{E}_g^{(d)}=-   \frac{1}{2\mu_d a_d^2}\label{8}  
\ee
 with $a_d$ real positive. The $\vp$ sum in Eq.~(\ref{7}) calculated with this $\mathcal{E}_g^{(d)}$ then gives
\be
\frac{1}{v^*}-\frac{1}{v}=\frac{1}{4\pi }\left(\frac{2\mu_d L^3}{a_d}\right)>0\,, \label{renorm_va}
\ee
which relates $v$ to $\mathcal{E}_g^{(d)}$ through $a_d$. As required, the obtained potential amplitude scales as the inverse of the sample volume, $L^3$.

The corresponding (normalized) dimer wave function, obtained from Eq.~(\ref{singlepaires1}), then reads 
\be
\lan \vp|\nu_g\ran= \sqrt{8\pi} \left(\frac{a_d}{L}\right)^{3/2} \frac{w_\vp}{1+(\vp a_d)^2}\,. \label{WFmomen}
\ee 
We see that $a_d$, which actually is the scattering length of the separable potential at hand --- indeed positive in the case of bound state --- is just the dimer spatial extension.

\section{Atom-dimer scattering\label{sec3:ad}}

We now consider the scattering between a $\alpha$ atom and a $(\alpha,\beta)$ dimer in a large sample.
The difference between the $(\alpha,\alpha,\beta)$ ground state energy $\mathcal{E}_g^{(ad)}$ and the one of the $(\alpha,\beta)$ dimer then  scales  as the inverse of the sample volume. The scattering length $a_{ad}$ between a $\alpha$ atom and a $(\alpha,\beta)$ dimer follows from this difference as 
\be
\Delta^{(ad)}=\mathcal{E}_g^{(ad)}-\mathcal{E}_g^{(d)} =4\pi\Big (\frac{a_{ad}}{2\mu_{ad} L^3}\Big)\,, \label{deltaT01}
\ee
where $\mu_{ad}^{-1} =m_\alpha^{-1}+(m_\alpha+m_\beta)^{-1}$ is the relative-motion mass for the atom-dimer pair.

To get this scattering length, we  have to determine the $(\alpha,\alpha,\beta)$ ground state energy, $(H-\mathcal{E}_g^{(ad)})|\Psi_g^{(ad)}\ran=0$.  
Since free $(\alpha,\beta)$ fermion pairs  can be written in terms of dimers according to
\be
B^\dag_{\vK,\vp}=\sum_i   B^\dag_i\,\,\delta_{\vK_i,\vK}\lan\nu_i |\vp\ran\, ,
\ee
as easy to check from Eq.~(\ref{BiBvKinu1}), we can expand the $(\alpha,\alpha,\beta)$ ground state on the (overcomplete) atom-dimer basis as
\be
|\Psi_g^{(ad)}\ran=\sum_{i,\vk}\Psi_{i,\vk}^{(ad)} B^\dag_i a^\dag_{\vk} |v\ran\,. \label{atomdimereigen}
\ee
Note that the commutation relation between the operators $B^\dag_i$ and $a^\dag_{\vk}$ reads
\be
0=\left[B^\dag_i, a^\dag_{\vk}\right]_\eta=B^\dag_i a^\dag_{\vk}+\eta\, a^\dag_{\vk} B^\dag_i\, ,\label{Biakccommu_eta} 
\ee
with $\eta=+$ when the operators $a^\dag_{\vk}$ and $b^\dag_{\vk^\prime}$ commute and $\eta=-$ when they anticommute,
\be
0=\left[b^\dag_{\vk'}, a^\dag_{\vk}\right]_{-\eta}\,,\label{bacpcommu_eta}
\ee
the value of the scattering length $a_{ad}$ being independent of $\eta$, as can be  ultimately checked.  

\begin{figure}[t!]
\begin{center}
\includegraphics[trim=2.5cm 4cm 1.8cm 4cm,clip,width=3.2in]{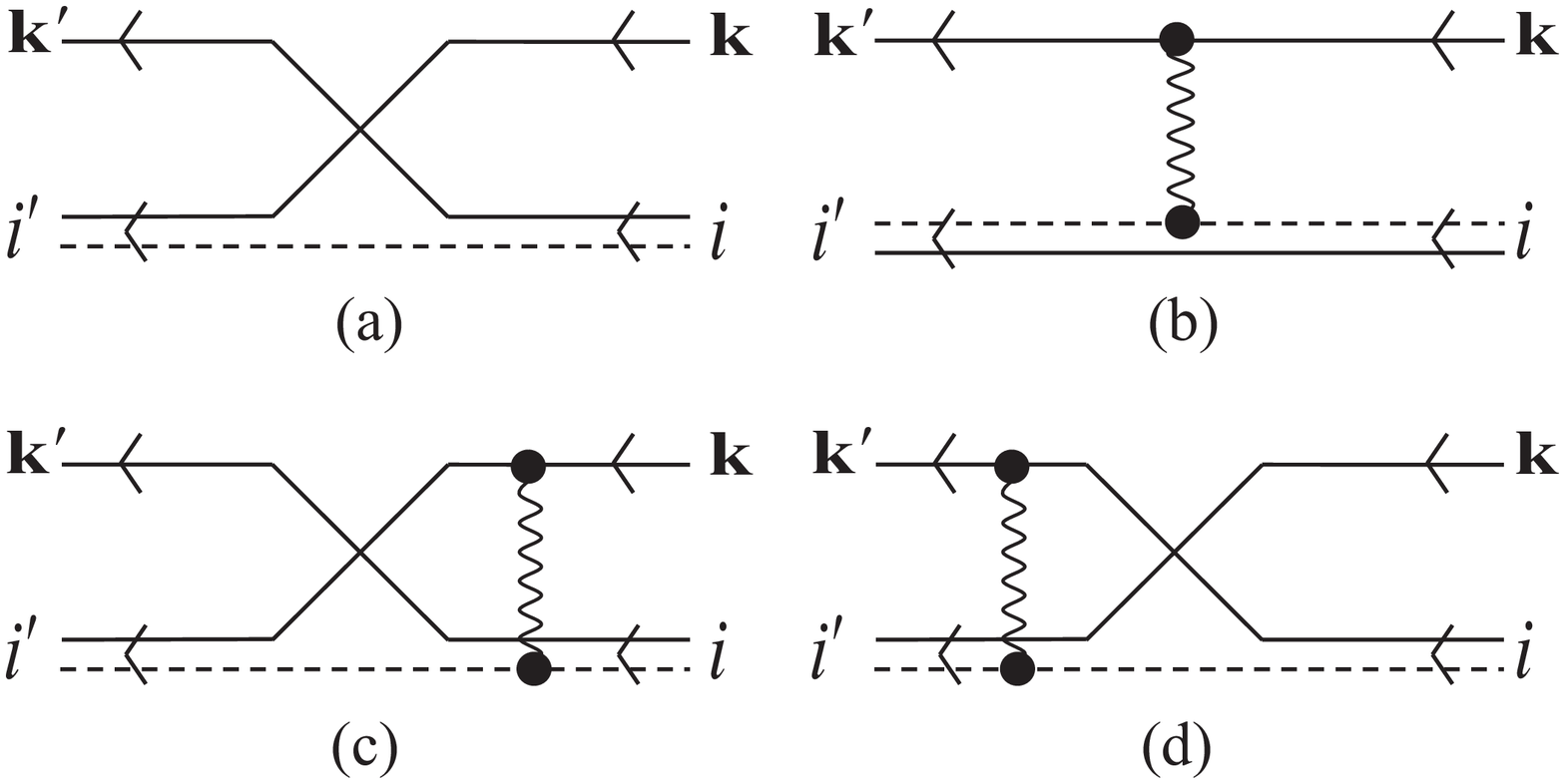}
\caption{\small Diagrammatic representations of the basic scatterings between a $\alpha$ atom and a dimer made of $(\alpha,\beta)$ atoms. (a) Pauli scattering $\lambda\big(_{i^\prime\, \,i}^{\vk^\prime\, \vk}\big)$, (b) direct interaction scattering $\xi\big(_{i^\prime\, \,i}^{\vk^\prime\, \vk}\big)$, (c) ``in" exchange interaction scattering $\xi^{in}\big(_{i^\prime\, \,i}^{\vk^\prime\, \vk}\big)$, (d) ``out" exchange interaction scattering $\xi^{out}\big(_{i^\prime\, \,i}^{\vk^\prime\, \vk}\big)$. Solid lines represent $\alpha$ atoms, dashed lines represent $\beta$ atoms, and double lines represent dimers made of $(\alpha,\beta)$ atoms.\label{fig:1} } 
   \end{center}
\end{figure}

\subsection{Coboson formalism}

To go further, we use  two commutation relations constructed along the coboson many-body formalism. The first one is a commutator (see \ref{app:sec1a})
\be
\left[H,B^\dag_i\right]_-
=E_i B^\dag_i+V_i^\dag\,\label{HBcommVi}\, .
\ee
It defines the creation potential $V^\dag_i$ that describes the interaction of the $(\alpha,\beta)$ dimer in state $i$ with the rest of the system. This operator gives zero when acting on vacuum, $V^\dag_i|v\ran=0$, while its interaction with a $\alpha$ atom follows from a second commutation relation (see \ref{app:sec1a}), which can be a commutator or an anticommutator depending on if the ($\alpha,\beta)$ fermion operators commute or anticommute, namely
\be
\left[V_i^\dag, a^\dag_{\vk}\right]_\eta
=\sum_{i^\prime,\vk^\prime}
\xi\big(_{i^\prime\, \,i}^{\vk^\prime\, \vk}\big)
 B^\dag_{i^\prime}a^\dag_{\vk^\prime}\,.\label{eq:Via}
\ee
The energy-like  scattering $\xi\big(_{i^\prime\, \,i}^{\vk^\prime\, \vk}\big)$ corresponds to the direct interaction of the $\alpha$ atom in state $\vk$ and the $\beta$ atom of the $(\alpha,\beta)$ dimer in state $i$, since $\alpha$ atoms interact with $\beta$ atoms only. It is precisely given by
\bea
\xi\big(_{i^\prime\, \,i}^{\vk^\prime\, \vk}\big)\!\!\!&=&\!\!\! 
{-}v\,\delta_{\vk^\prime{+}\vK_{i^\prime},\vk{+}\vK_i}\!\!
 \sum_{\vk_\alpha}
w_{ \gamma_\alpha (\vk_\alpha{-}\vK_{i^\prime}){+}\gamma_\beta\vk^\prime }w_{ \gamma_\alpha (\vk_\alpha{-}\vK_i){+}\gamma_\beta\vk }
\nn\\
&&\times\lan \nu_{i^\prime}|\vk_\alpha-\gamma_\alpha\vK_{i^\prime}\ran\lan\vk_\alpha-\gamma_\alpha\vK_i|\nu_i\ran\,, \label{17}
 \eea
  as can be directly read from its Shiva diagram  shown in Fig.~\ref{fig:1}(b) (see \ref{app:sec1b}). 
 

 These two commutators, used in the Schr\"{o}dinger equation $\big(H-\mathcal{E}_g^{(ad)}\big)|\Psi_g^{(ad)}\ran=0$, give
\be
0=\sum_{i,\vk}\Big\{\Big(E_{i,\vk}{-}\mathcal{E}_g^{(ad)}\Big) \Psi_{i,\vk}^{(ad)}{+}\sum_{i^\prime,\vk^\prime}\xi\big(_{i\, \,i^\prime}^{\vk\, \vk^\prime}\big) \Psi_{i^\prime,\vk^\prime}^{(ad)}  \Big\}  B^\dag_i a^\dag_{\vk}  |v\ran  \label{14}
\ee
with $E_{i,\vk }=E_i+\va_{\vk}^{(\alpha)}$.\

 In the next step, we project the above equation on the state $\lan v| a_{\vk^\prime} B_{i^\prime} $  and use the scalar product 
\be
\lan v| a_{\vk^\prime} B_{i^\prime}  B^\dag_i  a^\dag_{\vk} |v\ran=\delta_{i^\prime i}\delta_{\vk^\prime \vk}-\lambda\big(_{i^\prime\, \,i}^{\vk^\prime\, \vk}\big)\label{scalarproduaB}
\ee
that follows from two other commutators whatever $\eta$ (see \ref{app:sec1a})
\bea
\left[B_{i'},B^\dag_i\right]_-&=&\delta_{i' i}- D_{i'i}\,, \\
\left[D_{i'i}, a^\dag_{\vk}\right]_-&=&\sum_{\vk^\prime}\lambda\big(_{i^\prime\, \,i}^{\vk^\prime\, \vk}\big)  a^\dag_{\vk^\prime}\,.\label{Dipiavkdag} 
\eea
The dimensionless Pauli scattering  
$\lambda\big(_{i^\prime\, \,i}^{\vk^\prime\, \vk}\big)$ 
corresponds to fermion exchange between $\alpha$ atom and $(\alpha,\beta)$ dimer.
It is precisely given by
\be
\lambda\big(_{i^\prime\, \,i}^{\vk^\prime\, \vk}\big)\!=\!\delta_{\vk^\prime+\vK_{i^\prime},\vk+\vK_i}\lan \nu_{i^\prime}|\vk{-}\gamma_\alpha\vK_{i^\prime}\ran\lan \vk^\prime{-}\gamma_\alpha\vK_i|\nu_i\ran\,,\label{lambda:ipikpk}
\ee
as can be directly read from its Shiva diagram shown  in Fig.~\ref{fig:1}(a) (see \ref{app:sec1b}).\

 Equation (\ref{14}) then gives
\bd
0=\Big(E_{i,\vk}-\mathcal{E}_g^{(ad)}\Big)\Psi_{i,\vk}^{(ad)}+\sum_{ i^\prime, \vk^\prime }  \zeta \big(_{i\, \,i^\prime}^{\vk\, \vk^\prime}\big)\Psi_{i^\prime,\vk^\prime}^{(ad)}\,, \label{eq:Schrod_eX04}
\ed
where $\zeta\big(_{i^\prime\, \,i}^{\vk^\prime\, \vk}\big)$ 
is an effective scattering that contains direct and exchange processes (see \ref{app:sec1c})
\be
\zeta\big(_{i^\prime\, \,i}^{\vk^\prime\, \vk}\big)=\xi\big(_{i^\prime\, \,i}^{\vk^\prime\, \vk}\big)-\xi^{exch}_{int}\big(_{i^\prime\, \,i}^{\vk^\prime\, \vk}\big) -\xi^{exch}_{Pauli}\big(_{i^\prime\, \,i}^{\vk^\prime\, \vk}\big)\,. \label{zeta_ET} 
\ee
The exchange part of this scattering contains the expected ``in" and ``out" interaction scatterings (see Figs.~\ref{fig:1}(c,d)) with interaction between the ``in" and ``out" atom-dimer pairs, respectively,
\be
\xi^{exch}_{int}\big(_{i^\prime\, \,i}^{\vk^\prime\, \vk}\big)=\frac{\xi^{ in}\big(_{i^\prime\, \,i}^{\vk^\prime\, \vk}\big)+\xi^{ out}\big(_{i^\prime\, \,i}^{\vk^\prime\, \vk}\big)}{2}\, .\label{exchginterii}
\ee
The ``in" exchange interaction scattering, shown by the Shiva diagram of Fig.~\ref{fig:1}(c), is defined as $\xi^{in}\big(_{i^\prime\, \,i}^{\vk^\prime\, \vk}\big)=\sum_{j,\vp}\lambda \big(_{i^\prime\, \,j}^{\vk^\prime\, \vp}\big)\xi \big(_{j\, \,i}^{\vp\, \vk}\big)$.
With the help of Eq.~(\ref{singlepaires}), it can be written in a form such that the potential amplitude $v$ does not appear explicitly, namely (see \ref{app:sec1b})
\bea
\xi^{in}\big(_{i^\prime\, \,i}^{\vk^\prime\, \vk}\big)&=&
-\,\delta_{\vk^\prime+\vK_{i^\prime},\vk+\vK_i}\big(\va_{\vk-\gamma_\alpha\vK_{i'}}-\va_{\nu_{i'}}\big)\nn\\
&&\times\lan \nu_{i^\prime}|\vk-\gamma_\alpha\vK_{i^\prime}\ran
\lan \vk'-\gamma_\alpha\vK_{i}|\nu_i\ran\,, \label{xi_inad}
\eea
with $(\va_{\vk-\gamma_\alpha\vK_{i'}}-\va_{\nu_{i'}})$ replaced by $(\va_{\vk^\prime-\gamma_\alpha\vK_{i}}-\va_{\nu_{i}})$ in the case of $\xi^{out}\big(_{i^\prime\, \,i}^{\vk^\prime\, \vk}\big)$.

The $\zeta \big(_{i^\prime\, \,i}^{\vk^\prime\, \vk}\big)$ effective scattering also has a less obvious part, namely   the dimensionless  Pauli scattering multiplied by an energy that makes it energy-like,
\be
\xi^{exch}_{Pauli}\big(_{i^\prime\, \,i}^{\vk^\prime\, \vk}\big)=\lambda\big(_{i^\prime\, \,i}^{\vk^\prime\, \vk}\big)\Big(\frac{E_{i^\prime, \vk^\prime}+E_{i, \vk}}{2}-\mathcal{E}_g^{(ad)}\Big)\, .\label{exchgPauliii}
\ee
It is possible to show (see  \ref{app:sec1c}) that the effective scattering which rules the $(\alpha,\alpha,\beta)$ ground state fulfills $\zeta \big(_{i\, \,i^\prime}^{\vk\, \vk^\prime}\big)^*=\zeta \big(_{i^\prime\, \,i}^{\vk^\prime\, \vk}\big)$, 
as physically required by time reversal symmetry.\

 To obtain $\Delta^{(ad)}$ in Eq.~(\ref{deltaT01}), we  replace $\mathcal{E}_g^{(ad)}$ by $\mathcal{E}_g^{(d)}+\Delta^{(ad)}$ in Eq.~(\ref{eq:Schrod_eX04}), but by $\mathcal{E}_g^{(d)}$ only in Eq.~(\ref{exchgPauliii}) because $\Delta^{(ad)}$ scales as $1/L^3$, so it gives a contribution vanishingly small compared to the other terms of  $\zeta \big(_{i\, \,i^\prime}^{\vk\, \vk^\prime}\big) $.
 Next, in the sum of Eq.~(\ref{eq:Schrod_eX04}), we separate the lowest energy term
  $\{g,\v0\}$ from the other terms $\{i,\vk\}\not=\{g,\v0\}$. As $\mathcal{E}_g^{(d)}=E_{g,\v0}=\va_{\nu_g}$, the resulting coupled equations can be written in a compact form as
  \be
 \Big(\Delta^{(ad)}-\hat \zeta \big(_{g\, \,g}^{\v0\, \v0}\big)\Big) \Psi_{g,\v0}^{(ad)}=0\,,\label{27} 
 \ee
where the scattering $\hat \zeta$ is solution of the integral equation
\be
\hat \zeta \big(_{i\, \,g}^{\vk\, \v0}\big) = \zeta\big(_{i\, \,g}^{\vk\, \v0}\big)+\!\!\sum_{i',\vk'\not=g, \v0 } \!\! \zeta \big(_{i\, \,i^\prime}^{\vk\, \vk^\prime}\big)\frac{1}{\mathcal{E}_g^{(ad)}
{-}E_{i',\vk'}}\hat \zeta \big(_{i^\prime\, \,g}^{\vk^\prime\, \v0}\big)\,.\label{integralad}
\ee
In the above integral equation, we can replace $\mathcal{E}_g^{(ad)}=E_{g,\v0}+\Delta^{(ad)}$   by $E_{g,\v0}$ because
$\Delta^{(ad)}\propto1/L^3$ is negligible in front of any energy difference for $(i',\vk')\not=(g, \v0 )$. Since $\Psi_{g,\v0}^{(ad)}\not=0$ by construction, Eq.~(\ref{27}) readily gives $\Delta^{(ad)}=\hat \zeta \big(_{g\, \,g}^{\v0\, \v0}\big) $. \

Next, we relate the  scattering length $a_{ad}$ of an atom and a dimer in the large volume $L^3$ to the ground state energy $\mathcal{E}_g^{(ad)}$, as  in Eq.~(\ref{deltaT01}). To do it, we first rewrite Eq.~(\ref{eq:Schrod_eX04}) as a Lippmann- Schwinger-like equation
\bd
\Psi_{i,\vk}^{(ad)}=\delta_{i,g}\delta_{\vk,\v0}+ \frac{1}{\mathcal{E}_g^{(ad)}-E_{i,\vk}}\sum_{ i^\prime, \vk^\prime }  \zeta \big(_{i\, \,i^\prime}^{\vk\, \vk^\prime}\big)\Psi_{i^\prime,\vk^\prime}^{(ad)}\,. \label{eq:Schrod_eX0411}
\ed
We multiply the above equation by $\zeta \big(_{m\, i}^{\vp\, \vk}\big)$ and sum over $(i,\vk)$. This yields
\bd
\sum_{ i, \vk }  \zeta \big(_{m\, \,i}^{\vp\, \vk}\big)\Psi_{i,\vk}^{(ad)}=T_{m,\vp}=\zeta \big(_{m\, g}^{\vp\, \v0}\big)+\sum_{i,\vk}\frac{\zeta \big(_{m\, i}^{\vp\, \vk}\big)}{\mathcal{E}_g^{(ad)}-E_{i,\vk}}T_{i,\vk}\,.
\ed
By replacing $\mathcal{E}_g^{(ad)}$  with $E_{g,\v0}$ in the above equation and  using Eqs.~(\ref{27}) and (\ref{integralad}), we obtain $\Delta^{(ad)}=T_{g,0}$. Relating this quantity, which can be seen as a $T$-matrix element, to the scattering length of two particles allows us to obtain Eq.~(\ref{deltaT01}), or equivalently 
\be
a_{ad}=\frac{\mu_{ad}\,\,L^3}{2\pi} \hat \zeta\big(_{g\, \,g}^{\v0\, \v0}\big)\, .\label{defin:aad}
\ee 
The above expression of the atom-dimer scattering length is exact. The task now is to solve the integral equation (\ref{integralad}). This can be easily done by restricting the dimer relative-motion states in the $i'$ sum to the  ground state $\nu_g$, its energy and wave function reading in terms of the dimer scattering length $a_d$, as given in Eqs.~(\ref{8}) and (\ref{WFmomen}).

To go further, we note that the center-of-mass momentum of the $(g, \v0)$ pair is equal to zero, so are the ones of the $(i', \vk')$ pairs coupled to it in Eq.~(\ref{integralad}) because all scattering processes conserve momentum. As a result, the $\zeta \big(_{i\, \,i'}^{\vk\, \vk'}\big)$ scatterings of interest are $\zeta \big(_{(-\vk,\nu_g)\, (-\vk',\nu_g)}^{\hspace{0.3cm}\vk\,\,\hspace{1cm} \vk^\prime}\big)\equiv\zeta (\vk,\vk')$.
Next, we note that
\be
v=4\pi\left(\frac{ a_d}{2\mu_dL^3}\right)\frac{1}{(2q_c a_d/\pi)-1}\label{V_vq}
\ee
as obtained from Eqs.~(\ref{defv*}) and (\ref{renorm_va}). So, the potential amplitude $v$, and hence the direct interaction scattering $\xi$ (see Eq.~(\ref{17})),
go to zero when the cutoff $q_c$ gets large. By contrast, although the exchange interaction scattering also has a $v$ prefactor, the sum it contains  makes it  finite for large $q_c$, as seen from Eq.~(\ref{xi_inad}).  So, the direct interaction scattering can be neglected in front of the other parts of the $\zeta$ effective scattering (\ref{zeta_ET}).\

The relevant ``in" exchange interaction scattering, shown in Fig.~\ref{fig:1}(c) or \ref{fig:7}, precisely reads
\bea
\xi^{in}(\vk,\vk')&=&-(\va_{\vk'+\gamma_\alpha\vk}-\va_{\nu_g})\lan \nu_g|\vk'+\gamma_\alpha \vk\ran \lan \vk+\gamma_\alpha \vk'|\nu_g\ran\nn\\
&=&-8\pi\left(\frac{a_d}{2\mu_d L^3}\right)\frac{w_{\vk'+\gamma_\alpha \vk}w_{\vk+\gamma_\alpha \vk'}}{1+a_d^2 (\vk+\gamma_\alpha \vk')^2}\nn\\
&\equiv&-8\pi\left(\frac{a_d}{2\mu_d L^3}\right) I_{\bar{\vk},\bar{\vk'}}\,,\label{xiinkk'101}
\eea
with $\bar{\vk}=\vk a_d$,  the cutoff part being possibly replaced by 1 since $q_ca_d\gg1$. The ``out" exchange interaction scattering, shown in Fig.~\ref{fig:1}(d), reads as $\xi^{in}(\vk,\vk')$ with $I_{\bar{\vk},\bar{\vk'}}$ replaced by $I_{\bar{\vk'},\bar{\vk}}$. Turning to the Pauli part of the effective scattering, we find that it reads
\bea
\xi^{exch}_{Pauli}(\vk,\vk')&=&4\pi\left(\frac{a_d}{2\mu_{ad} L^3}\right)(\bar{\vk'}^2+\bar{\vk}^2) I_{\bar{\vk},\bar{\vk'}}I_{\bar{\vk'},\bar{\vk}}\label{xiexch_Pauli}\\
&=& 4\pi\left(\frac{a_d}{2\mu_{d} L^3}\right)\frac{\bar{\vk'}^2+\bar{\vk}^2}{\bar{\vk'}^2-\bar{\vk}^2}\big(I_{\bar{\vk},\bar{\vk'}}-I_{\bar{\vk'},\bar{\vk}}\big)\, ,\nn
\eea 
as $\mu_{ad}(1-\gamma_\alpha^2)=\mu_d$. All this leads to 
\be
\zeta(\vk,\vk')=4\pi\left(\frac{a_d}{2\mu_{d} L^3}\right)\frac{2\bar{\vk}^2}{\bar{\vk}^2-\bar{\vk'}^2} I_{\bar{\vk},\bar{\vk'}}+(\bar \vk\longleftrightarrow \bar \vk')\, .\label{zetavkvk'beforeave}
\ee

The next step is to perform the angular integration
\bea
\bar I( \bar \vk, \bar \vk')&=&\frac{1}{2}\int_{0}^{\pi} \sin\theta_{\vk\vk'}\,\,d \theta_{\vk\vk'} I_{\bar \vk,\bar \vk'} \nn\\
&=&\frac{1}{4 \gamma_\alpha \bar k \bar k' }\ln\left(\frac{1+(\bar k+\gamma_\alpha \bar k')^2}{1+(\bar k-\gamma_\alpha\bar  k')^2}\right)\, .\label{Ikkaver_bar}
\eea
Inserting this result in  Eq.~(\ref{zetavkvk'beforeave}) gives  the effective scattering $\zeta$  as
\bea
\bar \zeta( \vk, \vk')&\equiv&\frac{1}{2}\int_{0}^{\pi} \sin\theta_{\vk\vk'}\,\,d \theta_{\vk\vk'} \zeta( \vk, \vk')\label{barzetakkprime}\\
&=&\!\!2\pi\! \left(\!\frac{ a_d}{2\mu_d \gamma_\alpha  L^3}\!\right)\!\frac{1}{\bar k \bar k'}\!\frac{\bar k^2}{\bar k^2-\bar k'^2}\ln\left(\frac{1+(\bar k+\gamma_\alpha \bar k')^2}{1+(\bar k-\gamma_\alpha\bar  k')^2}\right)\nn\\
&&+(\bar k\longleftrightarrow \bar k') \, .\nonumber
\eea
Using the above expression, it becomes easy to compute $\hat \zeta\big(_{g\, \,g}^{\v0\, \v0}\big)$ for the atom-dimer scattering length, by solving
\be
\hat \zeta \big(_{(-\vk,\nu_g)\, \,g}^{\hspace{0.4cm}\vk\,\hspace{0.5cm} \v0}\big) = \zeta\big(_{(-\vk,\nu_g)\, \,g}^{\hspace{0.4cm}\vk\,\hspace{0.5cm} \v0}\big)+\sum_{\vk'\not=\v0 }\frac{ \bar \zeta( \vk, \vk')}{(-\vk'^2/2\mu_{ad})}\hat \zeta \big(_{(-\vk',\nu_g)\, \,g}^{\hspace{0.4cm}\vk'\,\hspace{0.5cm} \v0}\big)\,.\label{integralad2}
\ee
Before going further, we wish to stress that although the integral equation (\ref{integralad}) is formally exact, the fact that we have restricted the relative-motion states in the $i'$ sum to the ground state, and thus obtained the effective scattering $\bar \zeta( \vk, \vk')$ makes our integral equation (\ref{integralad2})  different  from  the one obtained using a field theoretical approach\cite{Alzettopra2010}. Yet,  the value of the scattering length $a_{ad}$ we obtain is in excellent agreement with the one obtained by the field theoretical approach for all mass ratios. To better grasp why this is so, let us first consider very heavy and very light $\beta$ atoms.

\subsection{$m_\beta/m_\alpha\rightarrow \infty$}
When the $\beta$ atom is very heavy, 
$ \mu_d \simeq m_\alpha \simeq \mu_{ad}$ and $\gamma_\alpha\simeq0$; so,  the $m_\beta$ mass disappears from the problem. In this limit, Eq.~(\ref{barzetakkprime})  reduces to
\be
\bar \zeta( \vk, \vk')\simeq 8 \pi\left(\frac{ a_d}{2 m_\alpha L^3}\right)\frac{1}{(1+\bar k^2)(1+\bar k'^2)}\, ,
\ee
which comes from exchange interaction scatterings only. It is worth noting that $\zeta( \vk, \vk')$ in Eq.~(\ref{zetavkvk'beforeave}) is equal to this quantity  without averaging the angular momentum with respect to the dimer and the $\alpha$ atom. This shows that within the expansion of dimer $s$-like internal-motion  states, only  $s$-wave scattering survives in this  limit. In general, because  two $\alpha$ atoms do not interact, and  orbit concentrically around the heavy $\beta$ atom as a fixed center, with a zero-range potential  $v(r)=v \delta(r)$, the scattering waves with angular momentum index $\ell \not=0$ disappear, since they vanish at $r=0$ where the potential is non-zero.  \

 The integral equation (\ref{integralad}) then appears as
\bea
\hat\zeta( \vk,\v0)&\simeq& 8\pi\left(\frac{ a_d}{2m_\alpha L^3}\right)\frac{1}{1+\bar k^2}\\
&&\times\Bigg\{1-\frac{1}{2\pi^2}\left(\frac{2m_\alpha L^3}{ a_d}\right)\int_0^\infty  \frac{d\bar k'\hat\zeta(\vk',\v0)}{1+\bar k'^2}\Bigg\}\, .\nn
\eea
It has an obvious solution  $\hat\zeta(\vk,\v0)\simeq4\pi (a_d/2m_\alpha L^3)/(1+\bar k^2) $, from which we readily get the atom-dimer scattering length as 
\be
a_{ad}\simeq a_d\,,
\ee 
which is the result obtained from the field theoretical approach (see Eq.~(8) in \cite{Alzettopra2010}).

\subsection{$m_\beta/ m_\alpha\rightarrow 0$}

When the $\alpha$ atom is very heavy, $ \mu_d \simeq m_\beta$, $\gamma_\alpha\simeq1$, and $\mu_{ad}\simeq m_\alpha/2$. So, in this limit, the two masses remain in the problem. To study the $\gamma$ dependence of $a_{ad}$, we  expand the logarithmic function in Eq.~(\ref{barzetakkprime}) up to first order in $\gamma=m_\beta/m_\alpha$,  
\bea
\lefteqn{\ln\left(\frac{1{+}(\bar k{+}\gamma_\alpha \bar k')^2}{1{+}(\bar k{-}\gamma_\alpha\bar  k')^2}\right)\simeq \ln\left(\frac{1+(\bar k+\bar k')^2}{1+(\bar k- \bar k')^2}\right)}\hspace{8cm}\\
+2\gamma \left(\frac{1+\bar k(\bar k+\bar k')}{1+(\bar k+ \bar k')^2}-\frac{1+\bar k(\bar k-\bar k')}{1+(\bar k- \bar k')^2}\right)\,.\nn
\eea
The integral equation (\ref{integralad}) then appears, since  $\hat\zeta(\vk,\v0)=\hat\zeta(-\vk,\v0)$, as
\be
2\pi \gamma  \bar k ~\hat\zeta(\vk,\v0)\!\simeq\!\frac{8\pi^2 a_d}{m_\alpha L^3}\frac{\bar k}{1{+}\bar k^2} 
+\int_{-\infty}^\infty\!\!  \frac{d\bar k'}{\bar k'}\ln\left(1{+}(\bar k{-} \bar k')^2\right)\hat\zeta(\vk',\v0)\, ,
\ee
in agreement with the result obtained from the field theoretical approach\cite{Alzettopra2010}. Solving the above integral equation yields
\be a_{ad}\simeq a_d\ln \frac{m_\alpha}{m_\beta}\, .\ee 

\begin{figure}[t!]
\centering
\subfigure[]{\label{fig:3a} \includegraphics[trim=1cm 0.1cm 1cm 0.2cm,clip,width=3in]  {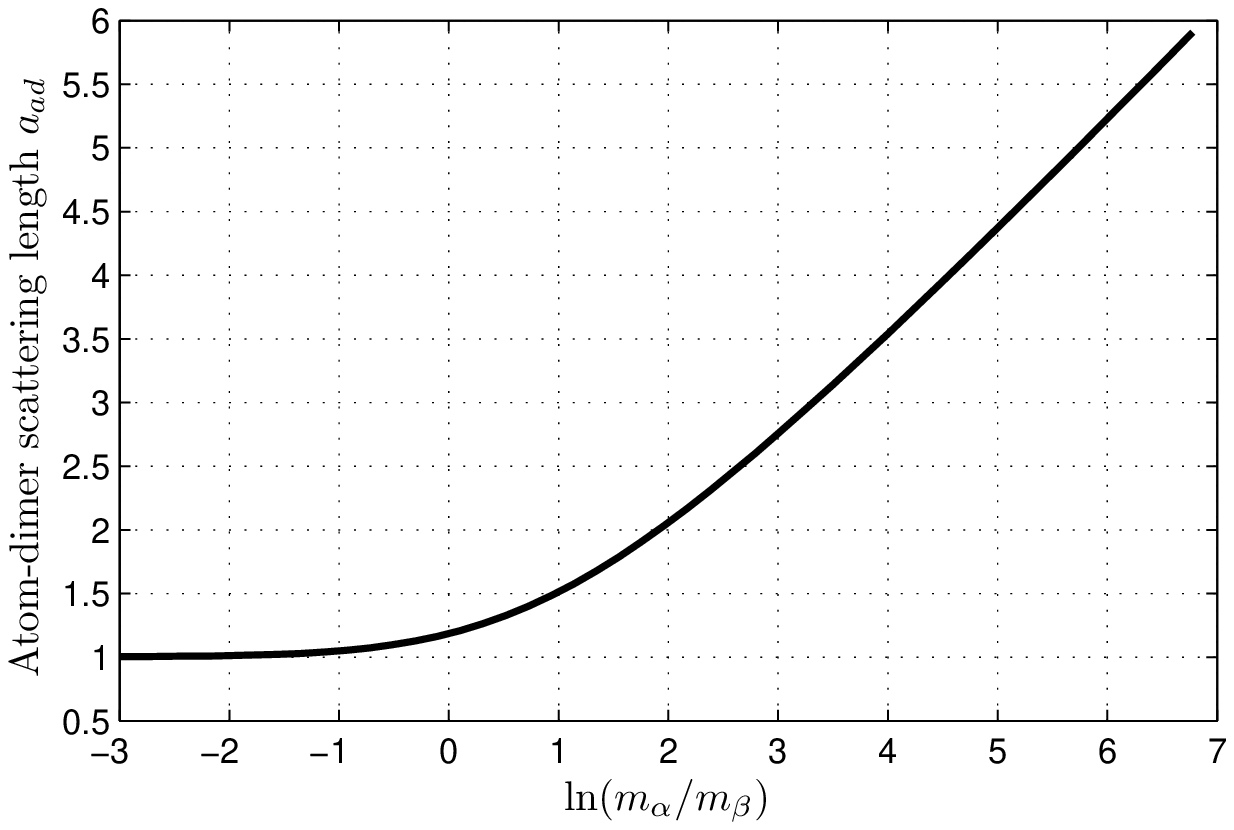}}
\subfigure[]{\label{fig:3b}  \includegraphics[trim=0.5cm 0.1cm 0.5cm 0.1cm,clip,width=3.3in]  {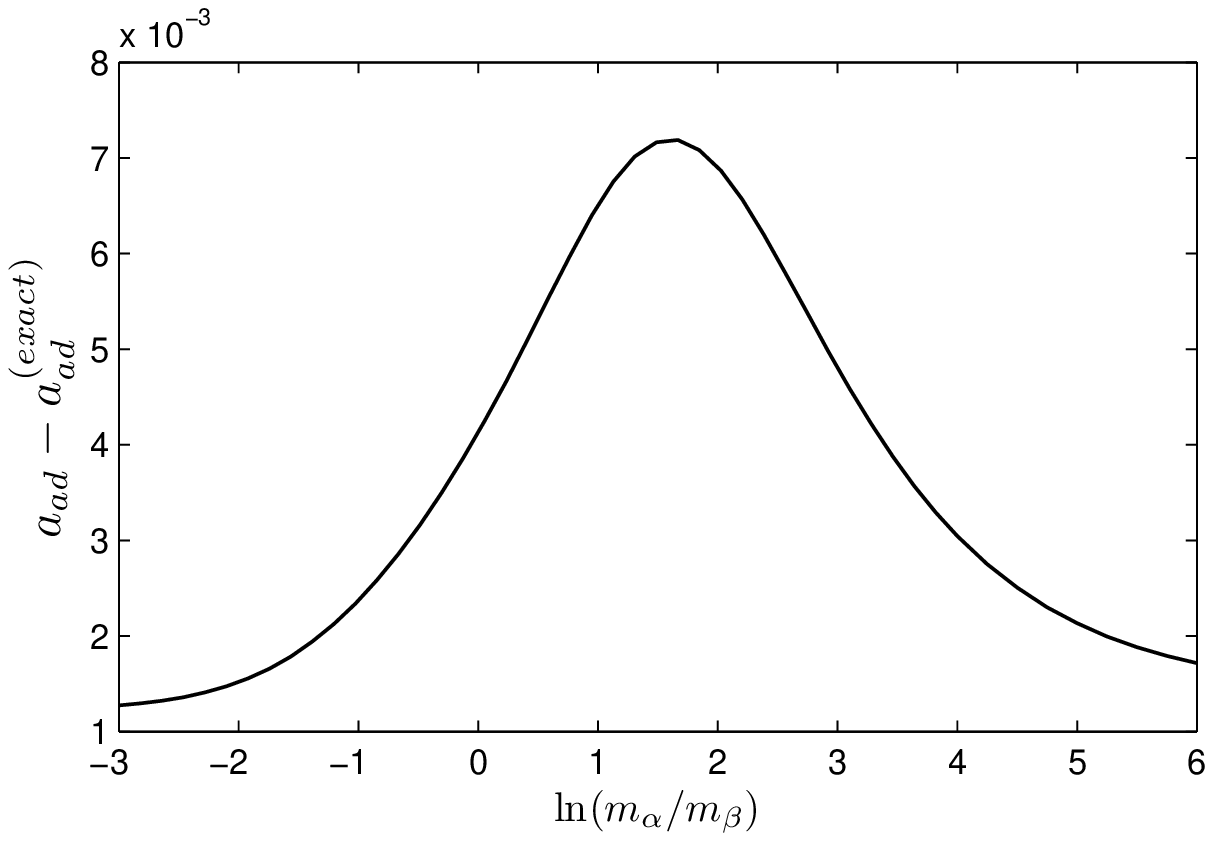}}
   \caption{\small (a) Atom-dimer scattering length (in units of $a_d$) obtained using Eq.~(\ref{barzetakkprime}). The value for $m_\alpha=m_\beta $ is $a_{ad}\simeq 1.179 a_d$. (b) The difference of our results with the exact values calculated with Eq.~(3) in \cite{Alzettopra2010}.  \label{fig:2} }
\end{figure}  

It is of interest to note that the  mass-dependent prefactor of the Pauli part (\ref{xiexch_Pauli}) in the effective scattering $\zeta$  is $1/\mu_{ad}$,  since the $(I_{\bar{\vk},\bar{\vk'}}-I_{\bar{\vk'},\bar{\vk}})$ difference scales as $1-\gamma_\alpha^2$. By contrast, the mass-dependent prefactor of the exchange interaction part (\ref{xiinkk'101}) is  $1/\mu_d$. So, in the large $m_\beta$  limit, these two parts of the effective scattering play an equally important role, while the scattering length in the small $m_\beta$ limit is controlled by  exchange interaction scatterings only.

\subsection{Numerical results}
The above results show that the integral equation we have derived  gives analytically exact atom-dimer scattering length in the large and small $m_\beta$ limits, despite the fact that we have obtained these limits with the dimer relative motion restricted to the ground state. For finite mass ratios, we have numerically solved Eq.~(\ref{integralad}) with the same effective scattering in Eq.~(\ref{barzetakkprime}). The results are shown in Fig.~\ref{fig:2}. Even for equal masses, the scattering length value is equal to 
\be
a_{ad}\simeq 1.183a_d\,,
\ee 
slightly larger than the exact value $1.179a_d$. The worst case is for $m_\alpha/m_\beta\sim 5$, which gives the discrepancy $0.007a_d$.

This leads us to conclude that our coboson many-body approach allows us to recover the previously obtained exact results for all mass ratios. 

The effectiveness of our restricting to the relative-motion ground state may come as a surprise at first. However, we must note that a short-range potential has one dimer bound state only, with an energy difference between the bound ground state and the other excited states large compared to the interaction scattering $\zeta$. So, terms with the intermediate dimer states in the integral equation (\ref{integralad}) taken from among unbound excited states are strongly suppressed by their energy denominator, instead of peaking at small momenta when the intermediate dimer state is the relative-motion ground state. While the contribution from the relative-motion excited states is physically expected to be small, it was far from clear that it indeed is so tiny.

\section{Dimer-dimer scattering\label{sec4:dd}}
\begin{figure}[t!]
\begin{center}
\includegraphics[trim=2.2cm 4cm 1.8cm 4cm,clip,width=3.2in]{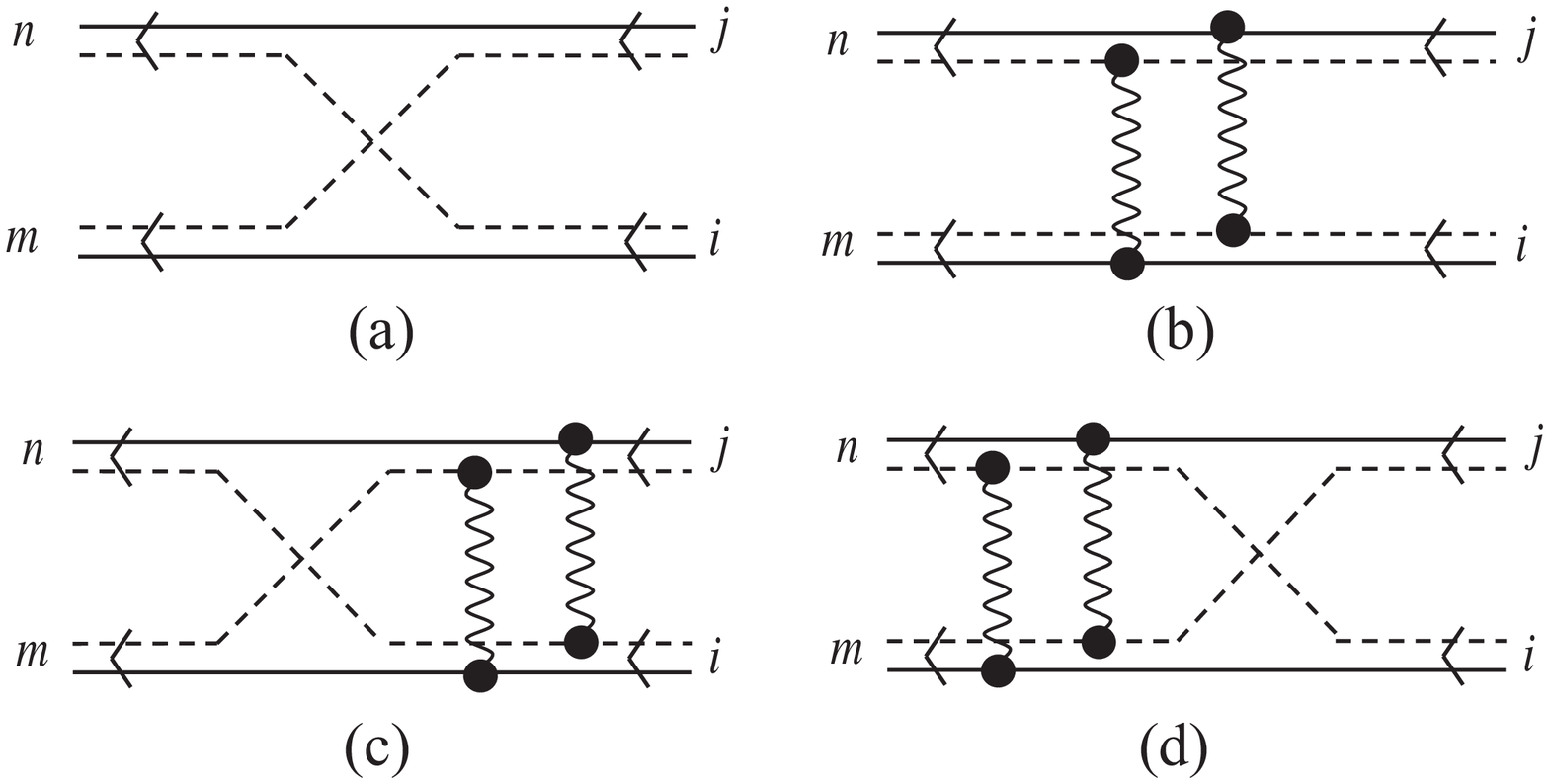}
\caption{\small Diagrammatic representations of various scatterings between dimer and dimer. (a) Pauli scattering (b) direct interaction scattering (c) ``in" exchange interaction scattering (d) ``out" exchange interaction scattering.  \label{fig:3}}
\end{center}
\end{figure}

In a previous work\cite{Shiau2016pra}, we have studied the dimer-dimer scattering length $a_{dd}$. It follows in a similar fashion from the ground state energy $\mathcal{E}_g^{(dd)}$ of $(\alpha,\alpha,\beta,\beta)$ fermions through 
\be
\Delta^{(dd)}=\mathcal{E}^{(dd)}_g-2\mathcal{E}_g^{(d)}= 4\pi\left( \frac{a_{dd}}{2 \mu_{dd} L^3}\right), \label{Delta}
\ee
with $\mu_{dd}^{-1}=2(m_\alpha+m_\beta)^{-1}$ being the relative-motion mass of the two dimers. As for atom-dimer scattering, the energy difference $\Delta^{(dd)}$ can be identified with $\hat{\zeta}\big(_{g\,g}^{g\,g}\big)$ obtained from the integral equation
\be
\hat{\zeta}\big(_{\,i\,g}^{j\,g}\big)=\zeta\big(_{\,i\,g}^{j\,g}\big)+\sum_{mn\neq gg}\zeta(_{im}^{ j\,n})\frac{1}{E_{gg}-E_{mn}}\hat{\zeta}\big(_{mg}^{n\,g}\big) \label{laddereq}
\ee
with $E_{ij}=E_i+E_j$. Here also, the kernel scattering 
\bea
            \zeta\big(_{mi}^{\, nj}\big)
    \!\!\!& =&\!\!\!\xi\big(_{mi}^{\, nj}\big){-}\frac{\xi^{in}\big(_{mi}^{\, nj}\big){+}\xi^{out}\big(_{mi}^{\, nj}\big)}{2} {-}\left(\frac{E_{mn}{+}E_{ij}}{2}{-}\mathcal{E}^{(dd)}_g\right)\lambda\big(_{mi}^{\, nj}\big)\nn\\
   &=&\xi\big(_{mi}^{\, nj}\big)-\xi^{exch}_{int}\big(_{mi}^{\, nj}\big)-\xi^{exch}_{Pauli}\big(_{mi}^{\, nj}\big)\label{eq:zeta1}
        \eea
contains all the fundamental dimer-dimer  scatterings,  represented by the Shiva diagrams of Fig.~\ref{fig:3}. Their precise expressions  are given in \ref{app:sec2}.    \

As for atom-dimer scattering, to obtain the dimer-dimer scattering length as a function of mass ratio, we have numerically computed Eq.~(\ref{laddereq}) with the dimer relative motion restricted to the ground state $\nu_g$ \cite{Shiau2016pra}. This amounts to neglecting  excitations to  relative-motion states other than the ground state. For the worst case, that is, for equal masses,  we obtained $0.64a_d$ instead of $0.60a_d$.  Although we have reported\cite{Shiau2016pra} the numerical value of this scattering length $a_{dd}$, it is of interest to analyze the limiting case when one of the atoms is very heavy, e.g., $ \gamma=m_\beta/m_\alpha\rightarrow 0 $, as we have done for atom-dimer scattering, in order to investigate the consequences of using a dimer basis restricted to  the relative-motion ground state.\

When the  $\alpha$ atom mass is large, $(\gamma_\beta,\gamma_\alpha)\rightarrow(0,1)$.  The ``in" exchange interaction scattering  given in Eq.~(\ref{app:inInt21})  and the Pauli scattering given in Eq.~(\ref{app:def_lambda}) reduce, for $\xi \big(_{\,\,( \vK',\nu_g )\,\,\,\,\,\,(\vK,\nu_g)}^{( -\vK', \nu_g)\,(-\vK,\nu_g)}\big)\equiv\xi( \vK',\vK)$, to 
\bea
\xi^{in}( \vK',\vK)&\simeq&-\frac{8\pi a_d}{\mu_d L^3}\frac{1 }{\bar \vQ^2+4}\,,\\
\lambda( \vK',\vK)&\simeq&\left(\frac{a_d}{L}\right)^3\frac{16\pi}{(\bar \vQ^2+4)^2}\,,
\eea
with $\bar \vQ=(\vK-\vK')a_d$. As for atom-dimer scattering, the direct interaction scattering $\xi$ given in Eq.~(\ref{app:def_dirXi}) contains a $v$ factor, which makes it negligible when  the  cutoff $q_c$ is large. Moreover,  for small $\gamma$, the  mass-dependent prefactor of the Pauli part, $\xi^{exch}_{Pauli}$, in the  $\zeta$ scattering is $\mu_d/\mu_{dd}$ ($\simeq\gamma$) smaller than the exchange interaction part, $\xi^{exch}_{int}$. So, in this limit, the dimer-dimer scattering is controlled by the exchange interaction part only. Equation (\ref{eq:zeta1}) then reduces to
\be
\zeta ( \vK',\vK)\simeq\frac{8\pi a_d}{\mu_d  L^3}\frac{1}{\bar \vQ^2+4} \,.
\ee

 After  angular averaging, we get 
\bea\bar \zeta(\vK',\vK)&=&\frac 1 2 \int^\pi_{0} \sin\theta_{\vK\vK'} d \theta_{\vK\vK'} \zeta( \vK',\vK)\nn\\
&\simeq& \frac{2\pi a_d}{\mu_d  L^3 \bar K \bar K'}\ln\left(\frac{4+(\bar K+\bar K')^2}{4+(\bar K-\bar K')^2}\right)\, .
\eea
Inserting this result into Eq.~(\ref{laddereq}) leads to
\bea
\lefteqn{\pi \gamma\bar K\hat\zeta(\vK,\v0)\simeq}\\
&&\frac{8\pi^2 a_d}{m_\alpha  L^3}\frac{\bar K}{4{+}\bar K^2}{-}\int^\infty_0\!\! \frac{d\bar K'}{\bar K'}\ln\!\left(\!\frac{4+(\bar K+\bar K')^2}{4{+}(\bar K{-}\bar K')^2}\!\right)\hat\zeta(\vK',\v0)\,.\nn
\eea
This equation is identical to the one obtained from a field theoretical approach (see Eq.~(56) in \cite{Alzetto2013}). Its  analytical resolution gives  the dimer-dimer scattering length as
\be
a_{dd}\simeq \frac{a_d}{2}\ln \frac{m_\alpha}{m_\beta}\simeq \frac{a_{ad}}{2}\, .
\ee 
We conclude that our coboson many-body procedure again gives the dimer-dimer scattering length very close to the exact value, except for a slight deviation near equal atom  masses.

\section{Conclusion}

We extend the coboson many-body formalism to cold atoms and derive the atom-dimer and dimer-dimer scattering lengths for arbitrary mass ratios. For atom-dimer, we recover the previously obtained  values for all masses. This remarkable agreement also holds true for dimer-dimer except near equal atom masses. These results show that restricting the dimer relative-motion basis to the ground state, as we have  done to numerically solve the integral equation for  effective scattering in an easy way, is an excellent approximation. Our work also shows that  atom-dimer scattering and  dimer-dimer scattering can be physically seen as a two-body scattering, provided fermion exchange with dimers is properly handled. 

Through Shiva diagrams, the coboson many-body formalism  provides a visual understanding of the role of the Pauli exclusion principle for composite bosons. Just like Feynman diagrams for elementary particles, these Shiva diagrams make the coboson formalism  an alternative powerful tool for understanding and calculating few-body as well as   many-body effects involving composite quantum particles, that is, nearly all particles of physical interest.

\section*{Acknowledgments}		
Y. C. C. and C. H. C acknowledge support from MOST 106-2112-M-001-022.
M.C. acknowledges many fruitful visits to Academia Sinica, Taiwan.

\renewcommand{\thesection}{\mbox{Appendix~\Roman{section}}} 
\setcounter{section}{0}
\renewcommand{\theequation}{\mbox{A.\arabic{equation}}} 
\setcounter{equation}{0} %
\section{Atom-dimer interaction \label{app:sec1}}

\subsection{Commutators or anticommutators\label{app:sec1a}}
The coboson formalism we have developed for fermion pairs interacting with fermion pairs makes use of commutators only. In its extension to dimer interacting with fermions, commutators are mixed with anticommutators when the $(\alpha,\beta)$ fermions commute,
\be
0=\left[b^\dag_{\vk'},a^\dag_\vk\right]_\eta=b^\dag_{\vk'}a^\dag_\vk+\eta a^\dag_\vk  b^\dag_{\vk'}\,,
\ee
with $\eta=(-)$ for a commutator and $\eta=(+)$ for an anticommutator. 
 Let us here show why this has to be so. The simplest way to understand when to use commutator or anticommutator for $(A,B)$ operators is to start with $AB$ and push $A$ to the right.\ 

(i) \textbf{The commutation relation between $(H,B^\dag_i)$ must be a commutator whatever $\eta$}

Let us consider $HB^\dag_i$. We  find, for ${[}a_{\vk'},a^\dag_\vk {]}_+=\delta_{\vk'\vk}={[}b_{\vk'},b^\dag_\vk {]}_+$, 
\bea
H^{(\alpha)}_0 B^\dag_i&=& \sum_\vk \va_\vk^{(\alpha)} a^\dag_\vk a_\vk \sum_\vp a^\dag_{\vp+\gamma_\alpha \vK_i}b^\dag_{-\vp+\gamma_\beta \vK_i}\lan \vp|\nu_i\ran\nn\\
&=& \sum_{\vk\vp} \va_\vk^{(\alpha)} a^\dag_\vk \left[a_\vk , a^\dag_{\vp+\gamma_\alpha \vK_i}\right]_+ b^\dag_{-\vp+\gamma_\beta \vK_i}\lan \vp|\nu_i\ran\nn\\
&&-\sum_{\vk\vp} \va_\vk^{(\alpha)} a^\dag_\vk a^\dag_{\vp+\gamma_\alpha \vK_i} a_\vk   b^\dag_{-\vp+\gamma_\beta \vK_i}\lan \vp|\nu_i\ran\, .
\eea
The second term on the right-hand side of the above equation is equal to $+B^\dag_i H^{(\alpha)}_0$, with a similar result for $H^{(\beta)}_0 B^\dag_i$ and $V B^\dag_i$. So, by combining these terms with the corresponding left-hand sides, we end with  a commutator ${[}H,B^\dag_i{]}_-$  whatever $\eta$.

(ii) \textbf{The commutation relation between $(V_i^\dag,a^\dag_\vk)$ depends on $\eta$}

From the definition of $V_i^\dag$ given in Eq.~(\ref{HBcommVi}), we get
\bea
V_i^\dag a^\dag_\vk&=&\left(-E_i B^\dag_i+\left[H,B^\dag_i\right]_-\right)a^\dag_\vk\nn\\
&=& -E_iB^\dag_i a^\dag_\vk +H  B^\dag_i a^\dag_\vk- B^\dag_i H a^\dag_\vk\, .\label{app:IaViak}
\eea
To go further, we note that
\be
H_0 a^\dag_\vk= \va_\vk^{(\alpha)} a^\dag_\vk +a^\dag_\vk H_0 \, ,
\ee
which leads us to consider $\big[H_0 ,a^\dag_\vk\big]_-$ whatever $\eta$; and similarly for $Va^\dag_\vk$. Equation (\ref{app:IaViak}) then reads, with the help of Eq.~(\ref{Biakccommu_eta}), 
\bea
V^\dag_i a^\dag_\vk&=&\eta E_i a^\dag_\vk B^\dag_i -\eta\left(\left[H,a^\dag_\vk\right]_-+ a^\dag_\vk H\right)B^\dag_i\nn\\
&&-B^\dag_i \left(\left[H,a^\dag_\vk\right]_-+ a^\dag_\vk H\right)\, ,
\eea
which can be written in a compact form as
\bea
V^\dag_i a^\dag_\vk&=&-\eta  a^\dag_\vk V^\dag_i -\left[B^\dag_i ,\left[H,a^\dag_\vk\right]_-\right]_\eta\, .
\eea
This leads us to consider $\big[V^\dag_i, a^\dag_\vk\big]_\eta$, that is, whether to use a commutator or an anticommutator depends on $\eta$.

(iii) \textbf{The commutation relation between $(B_{i^\prime},B^\dag_i)$ must be a commutator whatever $\eta$}

From the definition of $B^\dag_i$ given in Eq.~(\ref{BiBvKinu1}), we find 
\bea
B_{i^\prime} B^\dag_i&=& \sum_{\vp\vp^\prime}\lan \nu_{i^\prime}|\vp^\prime\ran \lan \vp|\nu_{i}\ran\\
&& \times b_{-\vp^\prime+\gamma_\beta \vK_{i^\prime}}a_{\vp^\prime+\gamma_\alpha \vK_{i^\prime}}
 a^\dag_{\vp+\gamma_\alpha \vK_i}b^\dag_{-\vp+\gamma_\beta \vK_i}\, ,\nn
 \eea
 which also reads
 \bea
B_{i^\prime} B^\dag_i&=&\sum_{\vp\vp^\prime}\lan \nu_{i^\prime}|\vp^\prime\ran \lan \vp|\nu_{i}\ran\label{app:BiBi'3} \\
&&\times \Big\{b_{-\vp^\prime+\gamma_\beta \vK_{i^\prime}}\left[a_{\vp^\prime+\gamma_\alpha \vK_{i^\prime}},
 a^\dag_{\vp+\gamma_\alpha \vK_i}\right]_+ b^\dag_{-\vp+\gamma_\beta \vK_i}\nn\\
 &&-b_{-\vp^\prime+\gamma_\beta \vK_{i^\prime}}  a^\dag_{\vp+\gamma_\alpha \vK_i} a_{\vp^\prime+\gamma_\alpha \vK_{i^\prime}}
 b^\dag_{-\vp+\gamma_\beta \vK_i}\Big\}\, .\nn
\eea
By writing the second term of the above curly bracket as
\be
 a^\dag_{\vp+\gamma_\alpha \vK_i} b_{-\vp^\prime+\gamma_\beta \vK_{i^\prime}}b^\dag_{-\vp+\gamma_\beta \vK_i}a_{\vp^\prime+\gamma_\alpha \vK_{i^\prime}}\,,
\ee
and by using
\bea
 b_{-\vp^\prime+\gamma_\beta \vK_{i^\prime}}b^\dag_{-\vp+\gamma_\beta \vK_i}&=&\left[b_{-\vp^\prime+\gamma_\beta \vK_{i^\prime}},b^\dag_{-\vp+\gamma_\beta \vK_i}\right]_+\nn\\
 && -b^\dag_{-\vp+\gamma_\beta \vK_i}b_{-\vp^\prime+\gamma_\beta \vK_{i^\prime}}\, ,
\eea
 the second term of Eq.~(\ref{app:BiBi'3}) leads to $+B^\dag_i B_{i^\prime} $. So, here again, we must consider  a commutator $\big[B_{i^\prime},B^\dag_i\big]_-$ whatever $\eta$.\

(iv) \textbf{The commutation relation between  $(D_{i^\prime i},a^\dag_\vk)$ must be a commutator whatever $\eta$}

From the definition of $D_{i^\prime i}$ given in Eq.~(\ref{Dipiavkdag}), we get
\bea
D_{i^\prime i} a^\dag_\vk&=& \left(\delta_{i^\prime i}-\left[B_{i^\prime},B^\dag_i\right]_-\right)a^\dag_\vk\nn\\
&=& \delta_{i^\prime i}a^\dag_\vk- B_{i^\prime} B^\dag_i a^\dag_\vk + B^\dag_i B_{i^\prime}a^\dag_\vk\, .\label{app:Di'iadagk}
\eea
To go further, we note that 
\bea
B_{i^\prime} a^\dag_\vk&=& \sum_{\vp^\prime} \lan \nu_{i^\prime}|\vp^\prime\ran b_{-\vp^\prime+\gamma_\beta \vK_{i^\prime}}a_{\vp^\prime+\gamma_\alpha \vK_{i^\prime}} a^\dag_\vk\\
&=&\sum_{\vp^\prime} \lan \nu_{i^\prime}|\vp^\prime\ran b_{-\vp^\prime+\gamma_\beta \vK_{i^\prime}}\Big\{\left[a_{\vp^\prime+\gamma_\alpha \vK_{i^\prime}}, a^\dag_\vk \right]_+\nn\\
&& - a^\dag_\vk a_{\vp^\prime+\gamma_\alpha \vK_{i^\prime}}  \Big\}\, .\nn
\eea
As the second term on the right-hand side of the above equation is equal to  $-\eta a^\dag_\vk B_{i^\prime}$, we are  led to consider  $\big[B_{i^\prime}, a^\dag_\vk\big]_\eta$. \

 When used into Eq.~(\ref{app:Di'iadagk}), this gives
 \bea
 D_{i^\prime i} a^\dag_\vk&=& \delta_{i^\prime i}a^\dag_\vk + B^\dag_i\left( \left[B_{i^\prime},a^\dag_\vk\right]_\eta-\eta a^\dag_\vk B_{i^\prime}\right)\nn\\
 &&+\eta  \left( \left[B_{i^\prime},a^\dag_\vk\right]_\eta-\eta a^\dag_\vk B_{i^\prime}\right)B^\dag_i\nn\\
 &=& a^\dag_\vk D_{i^\prime i}  +\left[B^\dag_i ,\left[B_{i^\prime},a^\dag_\vk\right]_\eta\right]_\eta\, ,
 \eea
which again leads us to consider the commutator $\big[D_{i^\prime i}, a^\dag_\vk\big]_-$ whatever $\eta$.

\renewcommand{\theequation}{\mbox{B.\arabic{equation}}} 
\setcounter{equation}{0}
\subsection{Relevant atom-dimer scatterings\label{app:sec1b}}

In this appendix, we derive the various $(\xi, \xi^{in}, \xi^{out}, \lambda)$ scatterings that enter the effective scattering $\zeta$ in the case of cold atoms interacting through different-species fermions only. We will do it by reading their expressions from the Shiva diagrams that represent them. The algebraic procedure to recover these expressions starting from their definitions in terms of commutation relations, is similar to the one used for excitons (see \cite{Shiauprb2012}). 

(i) \textbf{Pauli scattering for fermion exchange}

\begin{figure}[h!]
\begin{center}
\includegraphics[trim=6cm 7.5cm 6.5cm 7cm,clip,width=2.8in]{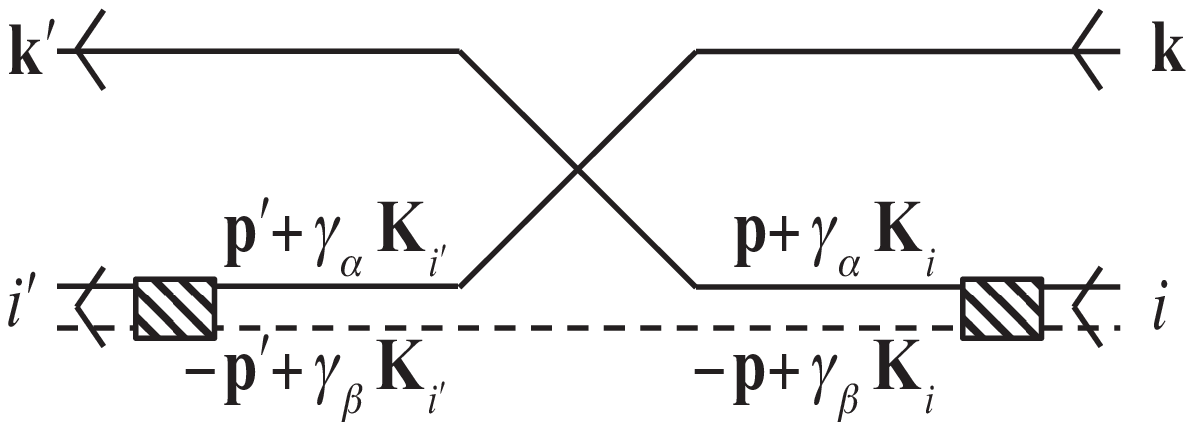}
\caption{\small Diagrammatic representation of the Pauli scattering $\lambda  \big(_{i^\prime\, \,i}^{\vk^\prime\, \vk}\big)$.  \label{fig:5}}
\end{center}
\end{figure}

The Pauli scattering $\lambda  \big(_{i^\prime\, \,i}^{\vk^\prime\, \vk}\big)$ is shown by the Shiva diagram of Fig.~\ref{fig:5}. The dimer $i$ expands on $B^\dag_{\vK_i,\vp}$ pairs through a vertex $\lan \vp|\nu_i\ran$, denoted by a dashed box. After exchanging with the $\alpha$ atom in state $\vk$, the $\alpha$ atom of the dimer $i$, initially in state $\vp+\gamma_\alpha \vK_i$, ends in state $\vk^\prime$; so, $\vk^\prime= \vp+\gamma_\alpha \vK_i$. Similarly, the $\alpha$ atom in state $\vk$ ends in state $\vp^\prime+\gamma_\alpha \vK_i^\prime$ to form the dimer $i^\prime$ through a vertex $\lan \nu_i^\prime|\vp^\prime\ran$; so, $\vk=\vp^\prime+\gamma_\alpha \vK_{i^\prime}$. In this exchange, the $\beta$ atom stays unchanged; so, $-\vp^\prime+\gamma_\beta \vK_{i^\prime}=-\vp+\gamma_\beta \vK_i$. All this leads to  
\bea
\lambda  \big(_{i^\prime\, \,i}^{\vk^\prime\, \vk}\big) &=&\sum_{\vp\vp^\prime}\lan \nu_{i^\prime}|\vp^\prime\ran\lan\vp|\nu_i\ran\label{app:lambda01a} \\
&&\times\delta_{\vk^\prime,\vp+\gamma_\alpha\vK_i}\delta_{\vk,\vp^\prime+\gamma_\alpha\vK_{i^\prime}}\delta_{-\vp^\prime+\gamma_\beta \vK_{i^\prime},-\vp+\gamma_\beta \vK_i}\,,\nn
\eea
which readily gives the expression of $\lambda  \big(_{i^\prime\, \,i}^{\vk^\prime\, \vk}\big)$ in Eq.~(\ref{lambda:ipikpk}).

(ii) \textbf{Direct interaction scattering}

\begin{figure}[h!]
\begin{center}
\includegraphics[trim=6cm 7.3cm 6.5cm 6.7cm,clip,width=2.8in]{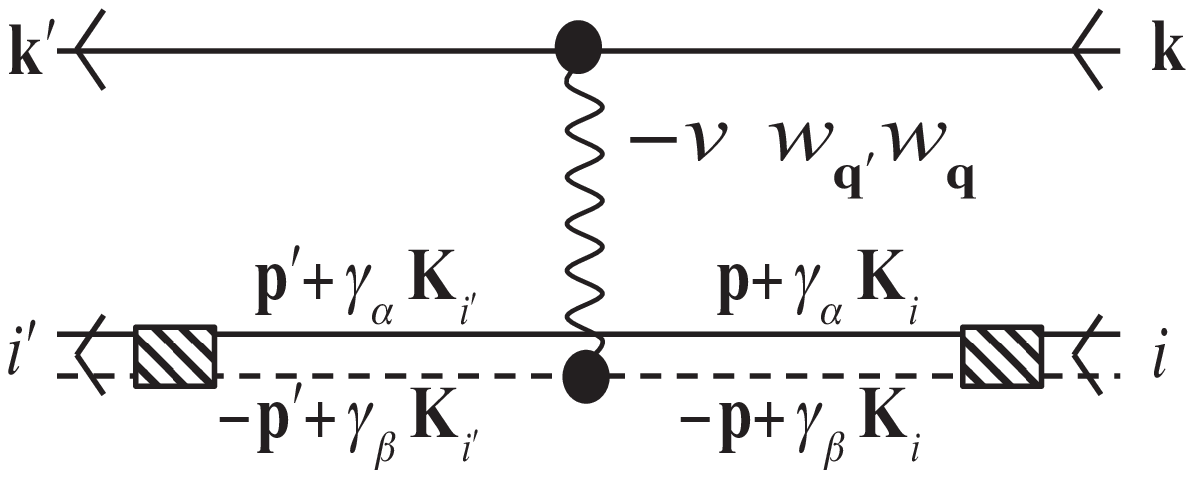}
\caption{\small Diagrammatic representation of the direct interaction scattering $\xi  \big(_{i^\prime\, \,i}^{\vk^\prime\, \vk}\big)$. \label{fig:6}}
\end{center}
\end{figure}

The direct interaction scattering $\xi  \big(_{i^\prime\, \,i}^{\vk^\prime\, \vk}\big)$ is shown by the Shiva diagram of Fig.~\ref{fig:6}. As for Pauli scattering, this scattering contains the two vertices $\lan \nu_i^\prime|\vp^\prime\ran$ and  $\lan \vp|\nu_i\ran$. Since $\alpha$ atoms do not interact between themselves, the $\alpha$ atom making the dimers $i$ and $i'$ stays unchanged; so, $\vk_\alpha=\vp^\prime+\gamma_\alpha \vK_{i^\prime}=\vp+\gamma_\alpha \vK_i$. By contrast, the $V$ potential transforms the pair made of the free $\alpha$ atom and the $\beta$ atom of the dimer $(\vk,-\vp+\gamma_\beta \vK_i)$ into $(\vk^\prime,-\vp^\prime+\gamma_\beta \vK_{i^\prime})$ with a scattering amplitude $-v w_{\vq^\prime} w_\vq$, where $\vq^\prime$ and $\vq$ are the final and initial relative-motion momenta of this pair, its center-of-mass momentum $\vK$  staying unchanged; so, $\vK=\vk-\vp+\gamma_\beta \vK_i=\vk^\prime-\vp^\prime+\gamma_\beta \vK_{i^\prime}$, while $\vk=\vq+\gamma_\alpha \vK$ and $\vk^\prime=\vq^\prime+\gamma_\alpha \vK$. All this leads to
\bea
\xi  \big(_{i^\prime\, \,i}^{\vk^\prime\, \vk}\big) &=&-v\sum_{\vp\vp^\prime}\lan \nu_{i^\prime}|\vp^\prime\ran\lan\vp |\nu_i\ran\sum_{\vk_\alpha}\delta_{\vk_\alpha,\vp+\gamma_\alpha\vK_i} \delta_{\vk_\alpha,\vp^\prime+\gamma_\alpha\vK_{i^\prime}}\nn\\
&&\times \sum_\vK \delta_{\vK,\vk^\prime-\vp^\prime+\gamma_\beta\vK_{i^\prime} } \delta_{\vK,\vk-\vp+\gamma_\beta\vK_i }\nn\\
&&\times \sum_{\vq\vq^\prime}w_\vq w_{\vq^\prime}\delta_{\vk,\vq+\gamma_\alpha \vK}\delta_{\vk^\prime,\vq^\prime+\gamma_\alpha \vK}\,,\,\label{app:direc01a}
\eea
which readily gives the expression of $\xi  \big(_{i^\prime\, \,i}^{\vk^\prime\, \vk}\big)$ in Eq.~(\ref{17}).

(iii) \textbf{``In" exchange interaction scattering}

\begin{figure}[h!]
\begin{center}
\includegraphics[trim=6.2cm 7.3cm 6cm 6.7cm,clip,width=2.8in]{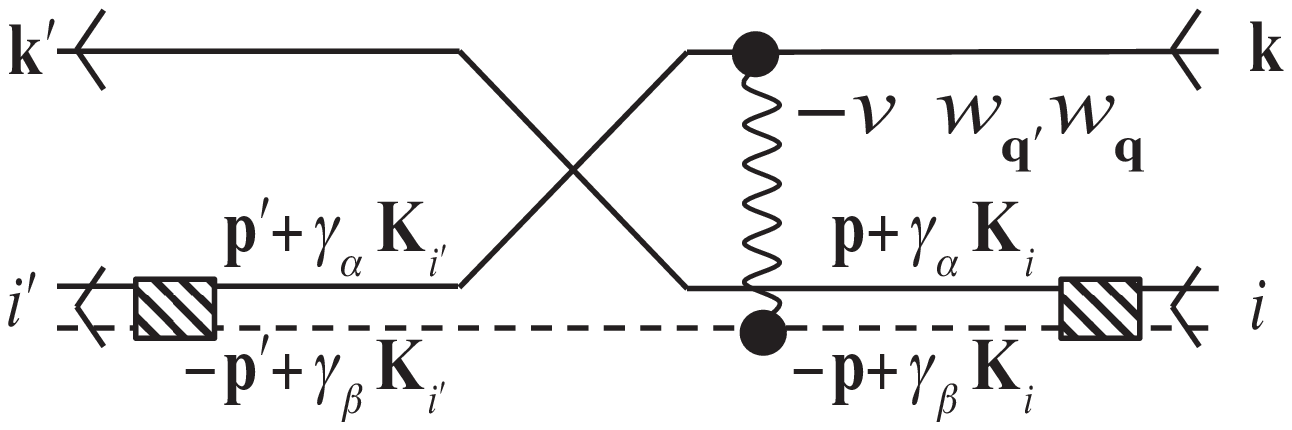}
\caption{\small Diagrammatic representation of the ``in" exchange interaction scattering $\xi^{in}  \big(_{i^\prime\, \,i}^{\vk^\prime\, \vk}\big)$. \label{fig:7}}
\end{center}
\end{figure}

The ``in" exchange interaction scattering $\xi^{in}  \big(_{i^\prime\, \,i}^{\vk^\prime\, \vk}\big)$ is shown by the Shiva diagram of Fig.~\ref{fig:7}. As for the direct interaction and Pauli scatterings, the exchange interaction scattering $\xi^{in}  \big(_{i^\prime\, \,i}^{\vk^\prime\, \vk}\big)$ contains the two vertices $\lan \vp|\nu_i\ran$ and $\lan \nu_i^\prime|\vp^\prime\ran$. Due to fermion exchange, the $\alpha$ atom  from the dimer  $i$ goes from state $\vp+\gamma_\alpha \vK_i$ to  state $\vk'$; so, $\vk'=\vp+\gamma_\alpha \vK_i$. The potential $V$ transforms the pair of $(\alpha,\beta)$ atoms $(\vk,-\vp+\gamma_\beta \vK_i)$ into the pair $(\vp'+\gamma_\alpha \vK_{i'},-\vp'+\gamma_\beta \vK_{i'})$ making the $i'$ dimer, with a scattering amplitude $-vw_{\vq'} w_\vq $ where $\vq'$ and $\vq$ are the  final and initial relative-motion momenta of the  pair, the center-of-mass momentum $\vK$ of this pair staying unchanged. So, $\vK=\vk-\vp+\gamma_\beta \vK_i =\vK_{i'}$, while $\vk=\vq+\gamma_\alpha \vK$ and $\vp'+\gamma_\alpha \vK_{i'}=\vq'+\gamma_\alpha \vK$. All this leads to
\bea
\xi^{in}  \big(_{i^\prime\, \,i}^{\vk^\prime\, \vk}\big) &=&-v\sum_{\vp\vp^\prime}\lan \nu_{i^\prime}|\vp^\prime\ran\lan\vp |\nu_i\ran\delta_{\vk',\vp+\gamma_\alpha\vK_i} \delta_{\vK_{i'},\vk-\vp+\gamma_\beta\vK_i}\nn\\
&&\times \sum_{\vq\vq^\prime}w_\vq w_{\vq^\prime}\delta_{\vk,\vq+\gamma_\alpha \vK_{i'}}\delta_{\vp^\prime,\vq^\prime}\,,\,\label{app:inexchg01a}
\eea
which readily gives
\bea
\xi^{in}  \big(_{i^\prime\, \,i}^{\vk^\prime\, \vk}\big) &=&-\delta_{\vk'+\vK_{i'},\vk+\vK_i}\lan \vk'-\gamma_\alpha\vK_i |\nu_i\ran\nn\\
&&\times\Big\{vw_{\vk-\gamma_\alpha \vK_{i'}}\sum_{\vp'}w_{\vp'} \lan \nu_{i^\prime}|\vp^\prime\ran\Big\}\,.\,\label{app:inexchg01b}
\eea
Actually,  it is possible to make the potential amplitude $v$  formally disappear from $\xi^{in}  \big(_{i^\prime\, \,i}^{\vk^\prime\, \vk}\big)$ by noting  that, through Eq.~(\ref{singlepaires}), the curly bracket of the above equation is equal to $(\va_{\vk-\gamma_\alpha\vK_{i'}}-\va_{\nu_{i'}})\lan \nu_{i'}|\vk-\gamma_\alpha\vK_{i'}\ran$.
So, we end with the expression of $\xi^{in}  \big(_{i^\prime\, \,i}^{\vk^\prime\, \vk}\big)$ given in Eq.~(\ref{xi_inad}).

\renewcommand{\theequation}{\mbox{C.\arabic{equation}}} 
\setcounter{equation}{0}
\subsection{Derivation of the effective scattering $\zeta \big(_{i\, \,i^\prime}^{\vk\, \vk^\prime}\big)$ \label{app:sec1c}}

In this appendix, we derive the effective scattering $\zeta \big(_{i^\prime\, \,i}^{\vk^\prime\, \vk}\big)$ 
in terms of the direct scattering $\xi$, the two  exchange interaction scatterings $(\xi^{in},\xi^{out})$, and the Pauli scattering $\lambda$.

We start with Eq.~(\ref{14}) that we project over the atom-dimer state $\lan v| a_{\vk'}B_{i'}$. With the help of Eq.~(\ref{scalarproduaB}), we find
\bea
0&=&\Big(E_{i',\vk'}-\mathcal{E}_g^{(ad)}\Big) \Psi_{i',\vk'}^{(ad)}+\sum_{i'',\vk''}\xi\big(_{i'\, \,i''}^{\vk'\, \vk''}\big) \Psi_{i'',\vk''}^{(ad)}  \nn\\ 
&&-\sum_{i,\vk}\lambda\big(_{i'\, \,i}^{\vk'\, \vk}\big)\Big(E_{i,\vk}-\mathcal{E}_g^{(ad)}\Big) \Psi_{i,\vk}^{(ad)}\nn\\
&&
-\sum_{i,\vk}\sum_{i'',\vk''}\lambda\big(_{i'\, \,i}^{\vk'\, \vk}\big)\xi\big(_{i\, \,i''}^{\vk\, \vk''}\big) \Psi_{i'',\vk''}^{(ad)}\, ,  \label{app:1c:14}
\eea
the scattering in the last term actually being the ``in" exchange interaction scattering
\be
\xi^{in}\big(_{i'\, \,i''}^{\vk'\, \vk''}\big) =\sum_{i,\vk}\lambda\big(_{i'\, \,i}^{\vk'\, \vk}\big)\xi\big(_{i\, \,i''}^{\vk\, \vk''}\big) \, .
\ee
As a result, Eq.~(\ref{app:1c:14}) is an integral equation for $\Psi_{i,\vk}^{(ad)}$,
\be
0=\Big(E_{i,\vk}{-}\mathcal{E}_g^{(ad)}\Big) \Psi_{i,\vk}^{(ad)}{+}\sum_{i',\vk'}\zeta\big(_{i\, \,i'}^{\vk\, \vk'}\big) \Psi_{i',\vk'}^{(ad)}\, ,
\ee
with an effective scattering $\zeta$ that appears as
\be
\zeta\big(_{i\, \,i'}^{\vk\, \vk'}\big) {=}\xi\big(_{i\, \,i'}^{\vk\, \vk'}\big) {-}\xi^{in}\big(_{i\, \,i'}^{\vk\, \vk'}\big) {-}\lambda\big(_{i\, \,i'}^{\vk\, \vk'}\big) \big(E_{i',\vk'}{-}\mathcal{E}_g^{(ad)}\big)\, .
\ee

We can write this effective scattering in a symmetric form with respect to  initial and final atom-dimer states by noting that 
\bd
\xi^{ in}\big(_{i\, \,i^\prime}^{\vk\, \vk^\prime}\big)+\lambda\big(_{i\, \,i^\prime}^{\vk\, \vk^\prime}\big)E_{i^\prime,\vk^\prime}=\xi^{ out}\big(_{i\, \,i^\prime}^{\vk\, \vk^\prime}\big)+ \lambda\big(_{i\, \,i^\prime}^{\vk\, \vk^\prime}\big) E_{i,\vk}\, ,\label{eq:rel_Pauliexchange}
\ed 
where the ``out" exchange interaction scattering is defined as $\xi^{out}\big(_{i\, \,i^\prime}^{\vk\, \vk^\prime}\big)=\sum_{i'',\vk''}\xi \big(_{i\, \,i''}^{\vk\, \vk''}\big)\lambda \big(_{i''\, \,i^\prime}^{\vk''\, \vk^\prime}\big) $.

A simple way to show the relation (\ref{eq:rel_Pauliexchange}) is to calculate $\lan v| a_\vk B_i H B^\dag_{i'}a^\dag_{\vk'}|v\ran$ using the commutator (\ref{HBcommVi}). We get, for $H$ acting on the right, 
\bea
\lan v| a_\vk B_i H B^\dag_{i'}a^\dag_{\vk'}|v\ran&=&E_{i^\prime,\vk^\prime}\lan v| a_\vk B_i B^\dag_{i'}a^\dag_{\vk'}|v\ran\\
&& + \sum_{i'',\vk''}\lan v| a_\vk B_i B^\dag_{i''}a^\dag_{\vk''}|v\ran \xi\big(_{i''\, \,i^\prime}^{\vk''\, \vk^\prime}\big)\, ,\nn
\eea
or with the help of Eq.~(\ref{scalarproduaB}), 
\be
E_{i,\vk}\delta_{ii'}\delta_{\vk\vk'}+\xi\big(_{i\, \,i^\prime}^{\vk\, \vk^\prime}\big)-\xi^{ in}\big(_{i\, \,i^\prime}^{\vk\, \vk^\prime}\big)-\lambda\big(_{i\, \,i^\prime}^{\vk\, \vk^\prime}\big)E_{i^\prime,\vk^\prime}\, .
\ee
We can also calculate this quantity via $\lan v| a_\vk B_i H B^\dag_{i'}a^\dag_{\vk'}|v\ran=\lan v| a_{\vk'} B_{i'} H B^\dag_{i}a^\dag_{\vk}|v\ran^*$ using the above equation. Equation (\ref{eq:rel_Pauliexchange}) then follows from $\xi\big(_{i^\prime\, \,i}^{\vk^\prime\, \vk}\big)^*=\xi\big(_{i\, \,i^\prime}^{\vk\, \vk^\prime}\big)$ and $\lambda\big(_{i^\prime\, \,i}^{\vk^\prime\, \vk}\big)^*=\lambda\big(_{i\, \,i^\prime}^{\vk\, \vk^\prime}\big)$, but $\xi^{in}\big(_{i^\prime\, \,i}^{\vk^\prime\, \vk}\big)^*=\xi^{out}\big(_{i\, \,i^\prime}^{\vk\, \vk^\prime}\big)$.

With the help of Eq.~(\ref{eq:rel_Pauliexchange}), it is easy to obtain the symmetric form given in Eq.~(\ref{zeta_ET}), which preserves the time reversal symmetry of the effective scattering, namely
\be\zeta\big(_{i\, \,i^\prime}^{\vk\, \vk^\prime}\big)^*=\zeta\big(_{i^\prime\, \,i}^{\vk^\prime\, \vk}\big)\, .
\ee


\section{Relevant dimer-dimer scatterings \label{app:sec2}}
Here, we recall the  $\xi$, $\xi^{out}$, and $\lambda$ scatterings  in the case of cold atoms. Detailed derivations in the case of excitons can be found in \cite{SYAnnals}. The direct interaction scattering between  $(\alpha_1,\beta_1)$ dimer and $(\alpha_2,\beta_2)$ dimer with the total center-of-mass momentum equal to zero  reads as
\bea
\xi\big(_{\,( \nu_m,\vK' )\,\,\,\,\,\,\,(\nu_i,\vK)}^{(  \nu_n,-\vK')\,(\nu_j,-\vK)}\big) =-v_{\vQ}\sum_{\vk,\vp}\Big(\lan \nu_m|\vk+\gamma_\beta \vQ\ran\lan\nu_n|\vp+\gamma_\alpha \vQ\ran\nn\\
+\lan \nu_m|\vk-\gamma_\alpha \vQ\ran\lan\nu_n|\vp-\gamma_\beta \vQ\ran\Big)\lan \vk|\nu_i\ran\lan \vp|\nu_j\ran\,,\label{app:def_dirXi}\hspace{1cm}
\eea
with $\vQ=\vK-\vK^\prime$, while the ``in" exchange interaction scattering is given by 
\bea
\xi^{ in}\big(_{\,( \nu_m,\vK' )\,\,\,\,\,\,\,(\nu_i,\vK)}^{(  \nu_n,-\vK')\,(\nu_j,-\vK)}\big)
\!\!&=&\!\!{-}\sum_{\vk,\vp}v_\vp\Big(\langle \nu_m|\vk{+}\frac{\vP_-}{2}{-}\vp\rangle\langle \nu_n|\vk{-}\frac{\vP_-}{2}\rangle\nn\\
&&+\langle \nu_m|\vk{+}\frac{\vP_-}{2}\rangle\langle \nu_n|\vk{-}\frac{\vP_-}{2}{-}\vp\rangle\Big)\nn\\
&&\times \langle \vk{+}\frac{\vP_+}{2}|\nu_i\rangle\langle \vk{-}\frac{\vP_+}{2}|\nu_j\rangle\,,\label{app:inInt2}
\eea
with  $\vP_\pm=\gamma_\beta (\vK'+\vK)\pm\gamma_\alpha(\vK^\prime-\vK)$.\ 

As for the atom-dimer scattering, we can eliminate the potential amplitude $v$ in this exchange interaction scattering  by  using Eq.~(\ref{singlepaires}). We then find
\bea
\xi^{in}\big(_{\,( \nu_m,\vK' )\,\,\,\,\,\,\,(\nu_i,\vK)}^{(  \nu_n,-\vK')\,(\nu_j,-\vK)}\big)
{=}\sum_{\vk}\left(\va_{\nu_m}{+}\va_{\nu_n}{-}\va_{\vk+ \vP_-/2}{-}\va_{\vk- \vP_-/2}\right)\nn\\
\times\langle \nu_m|\vk{+}\frac{\vP_-}{2}\rangle \langle \nu_n|\vk{-}\frac{\vP_-}{2}\rangle\langle \vk{+}\frac{\vP_+}{2}|\nu_i\rangle\langle \vk{-}\frac{\vP_+}{2}|\nu_j\rangle.\label{app:inInt21}\hspace{0.5cm}
\eea

Finally, the Pauli scattering for fermion exchange between two dimers is given by
\bea
\lefteqn{\lambda\big(_{\,( \nu_m,\vK' )\,\,\,\,\,\,\,(\nu_i,\vK)}^{(  \nu_n,-\vK')\,(\nu_j,-\vK)}\big)
=}\label{app:def_lambda}\\
&&\sum_\vk\langle \nu_m|\vk+\frac{\vP_-}{2}\rangle\langle \nu_n|\vk-\frac{\vP_-}{2}\rangle\langle \vk+\frac{\vP_+}{2}|\nu_i\rangle\langle \vk-\frac{\vP_+}{2}|\nu_j\rangle\,.\nn
\eea

\end{document}